\documentclass[a4paper,11pt]{article}

\usepackage[dvipdfmx]{graphicx}
\usepackage{natbib}
\usepackage{here}
\usepackage{setspace}
\usepackage{multirow}
\usepackage{amsmath, amssymb}
\usepackage{bm}
\usepackage{subfigure}

\newcommand{\str}[1]{\renewcommand{\baselinestretch}{#1}\normalsize}
\newcommand{\normalstr}{\str{1.3}}

\begin{document}

\str{1.05}

\title
{Bayesian modeling of temporal dependence in large sparse contingency tables}

\author{\textsc{Tsuyoshi Kunihama and David B. Dunson} \\
\textit{\small Department of Statistical Science, Duke University, Durham, NC 27708-0251, USA} \\
\texttt{\small tsuyoshi.kunihama@stat.duke.edu} \\
\texttt{\small dunson@stat.duke.edu} 
}

\date{May, 2012}

\maketitle

\vspace{-8mm}


\begin{abstract}

\noindent
In many applications, it is of interest to study trends over time in relationships among categorical variables, such as age group, ethnicity, religious affiliation, political party and preference for particular policies.  At each time point, a sample of individuals provide responses to a set of questions, with different individuals sampled at each time.  In such settings, there tends to be abundant missing data and the variables being measured may change over time.  At each time point, one obtains a large sparse contingency table, with the number of cells often much larger than the number of individuals being surveyed.  To borrow information across time in modeling large sparse contingency tables, we propose a Bayesian autoregressive tensor factorization approach.  The proposed model relies on a probabilistic Parafac factorization of the joint pmf characterizing the categorical data distribution at each time point, with autocorrelation included across times.  Efficient computational methods are developed relying on MCMC.   The methods are evaluated through simulation examples and applied to social survey data.  
\\

\noindent
\textit{Key words:} Dynamic model; Multivariate categorical data; Nonparametric Bayes; Panel data; Parafac; Probabilistic tensor factorization; Stick-breaking.

\end{abstract}



\str{1.55}

\section{Introduction}

Time-indexed multivariate categorical data are collected in many areas, with partially-overlapping categorical variables measured for different subjects at the different time points.  As a motivating application, we consider social science surveys that are conducted at regular time intervals, containing many categorical questions such as gender, race, age group, ethnicity, religious affiliation, political party and preference for particular policies.  For such surveys and other types of time-indexed multivariate categorical data, it is common for the variables measured (questions asked) to vary somewhat over time while a subset of the variables will be measured at all times.  In addition, the number of variables measured can be moderate to large leading to a contingency table with an {\em enormous} number of cells, the vast majority of which are empty.  Given the fact that social science data often contain complex interactions, it becomes extremely challenging to build realistic and computationally tractable models that allow ultra-sparse data.  We define ultra-sparse contingency tables as having exponentially or super-exponentially more cells than the sample size.

Let ${\bf x}_{ti} = (x_{ti1},\ldots,x_{tip})'$ denote the multivariate response for the $i$th subject in the survey at time $t$, with the $j$th categorical question having $d_j$ elements, $x_{tij} \in \{1,\ldots,d_j\}, j=1,\ldots, p$.  We accommodate the case in which the specific variables measured can vary across time by introducing missingness indicators, $m_{ti} = (m_{ti1},\ldots, m_{tip})'$, with $m_{tij} = 1$ if variable $j$ is missing for subject $i$ at time $t$; we allow design-based missingness in which certain variables are not measured for any subjects at a particular time and for individual-specific missingness in which certain individuals fail to answer all the questions posed to them.  In both cases we assume missing at random.

There is a rich literature on the analysis of contingency tables (\cite{Agresti02}; \cite{FienbergRinaldo07}).  Log linear models are perhaps the most commonly used modeling framework.  Routine implementations rely on maximum likelihood estimation, though there is also a rich Bayesian literature.  For large, sparse contingency tables, maximum likelihood estimates do not exist in many cases except for overly-simplistic log-linear models and richer classes of models become challenging to implement computationally.  There is a rich literature on graphical modeling approaches to estimating conditional independence structures in categorical variables, with \cite{DobraLenkoski11} proposing a recent Bayesian approach. 
Although their method is computationally efficient, except for very small tables, the number of possible graphical models is so enormous that is becomes infeasible to visit more than a vanishingly small fraction of the models making accurate model selection or averaging difficult.  To facilitate scaling to large tables, \cite{DunsonXing09} and \cite{BhattacharyaDunson11} recently proposed Bayesian probabilistic tensor factorizations.  These methods express the probability tensor corresponding to the joint probability mass function of the categorical variables as a convex combination of independent components.  Such methods have not yet been developed for time-indexed contingency tables.

There is a rich literature on categorical time series and longitudinal data analysis in which the same categorical variable is repeatedly measured for each subject over time.  For example, Markov models, state space models and random effects models are routinely applied in such settings.  However, these models are not relevant to the problem of incorporating dependence over time in modeling of large sparse contingency tables.  As different subjects are measured at different times, we are not faced with the problem of incorporating within-subject dependence in repeated observations; instead our goal is to include dependence in the parameters characterizing the time-dependent joint pmfs for the categorical variables.  To our knowledge, this problem has not yet been addressed in the literature.  Although one can potentially adapt log-linear or graphical models developed for a contingency tables at one time in a somewhat straightforward manner, the hurdles mentioned above for the static case are compounded in the dynamic setting.

To facilitate routine implementations in ultra sparse cases, we propose to adapt the Dunson and Xing (DX) (2009) probabilistic Parafac factorization to the dynamic setting.  The DX model induces a tensor factorization through a Dirichlet process (DP) mixture of product multinomial distributions for the categorical observations.  There is an increasingly rich literature proposing nonparametric Bayes dynamic models, which allow time-indexed dependent random probability measures.  Perhaps the most common approach relies on a dependent DP (\cite{MacEachern99, MacEachern00}), which incorporates time dependence in the weights and/or atoms in a stick-breaking representation (\cite{GriffinSteel06}; \cite{RodriguezHorst08}; \cite{ChungDunson11}).  Most applications of dependent DPs fix the weights and allow the atoms to vary, as varying weights can lead to computational complexities.  For dynamic modeling of contingency tables, it is more parsimonious to allow varying weights and varying atoms can lead to a substantial computational burden.  An alternative approach, which allows varying weights in a computationally convenient and flexible manner, relies of dynamic mixtures of DPs (\cite{Dunson06}; \cite{RenDunsonLindrothCarin10}).  Recently, a class of probit stick-breaking processes was proposed (\cite{ChungDunson09}), which has the appealing feature of allowing one to induce time dependence in random probability measures through Gaussian time series models (\cite{RodriguezDunson11}).

We propose a new nonparametric state space model for time-indexed ultra sparse contingency tables.  Relying on a DX-type probabilistic Parafac factorization, we place a dynamic model on the weights, which relies on transformed normal random variables in a similar manner to probit stick-breaking.  The model is nonparametric in the sense that the induced prior for each time-indexed joint pmf assigns positive probability in arbitrarily small neighborhoods of any ``true'' data-generating pmf.  Hence, our model can allow higher-order interactions and complex dependences, while shrinking towards a low-dimensional structure and borrowing information across time to address the curse of dimensionality.  In addition, and crucially for the approach to be useful in the motivating applications, posterior computation can be implemented via a highly efficient Markov chain Monte Carlo (MCMC) algorithm relying on a slice sampler related to \cite{KalliGriffinWalker11}.  Finally, the factorization produces a low-dimensional representation of the joint pmf, which is otherwise characterized by a daunting number of parameters in many cases, as the number of cells of the tables can be truly massive.

\normalstr
\section{Model specification}

\subsection{Modeling of multivariate categorical data}

We review the nonparametric Bayes approach of \cite{DunsonXing09} for a static large sparse contingency table.  Let ${\bf x}_i=(x_{i1},\ldots,x_{ip})'$ be multivariate categorical data for the $i$th subject, with $x_{ij} \in \{1,\ldots, d_j \}$, $j=1,\ldots, p$.  Let
\begin{align*}
\bm{\pi} = \left\{ \pi_{c_1\cdots c_p}, \, c_j=1,\ldots,d_j, \, j=1,\ldots,p \right\} \in \Pi_{d_1\cdots d_p}
\end{align*}
be a probability tensor where $\pi_{c_1\cdots c_p}=P(x_{i1}=c_1,\ldots,x_{ip}=c_p)$ is a cell probability and $\Pi_{d_1\cdots d_p}$ is the set of all probability tensors of size $d_1\times \cdots \times d_p$.  \cite{DunsonXing09} show that any $\bm{\pi} \in \Pi_{d_1\cdots d_p}$ can be decomposed as
\begin{align}
\bm{\pi} &= \sum_{h=1}^{k} \nu_{h} \Psi_h, \ \ \ \ \Psi_h = \bm{\psi}_h^{(1)} \otimes \cdots \otimes \bm{\psi}_h^{(p)}
\label{eq:mix}
\end{align}
where $\bm{\nu}=(\nu_1,\ldots,\nu_k)'$ is a probability vector, $\Psi_h\in \Pi_{d_1\cdots d_p}$ and $\bm{\psi}_h^{(j)}=(\psi^{(j)}_{h1},\ldots,\psi^{(j)}_{hd_j})'$ is a $d_j\times 1$ probability vector for $h=1,\ldots,k$ and $j=1,\ldots,p$.  This expression relies on a Parafac tensor factorization (\cite{Harshman70} and \cite{Kolda01}).  It follows that any multivariate categorical data distribution can be expressed as a mixture of product multinomials,
\begin{align*}
P(x_{i1}=c_1,\ldots,x_{ip}=c_p) = \pi_{c_1\cdots c_p} = \sum_{h=1}^{k} \nu_{h} \prod_{j=1}^p \psi_{hc_j}^{(j)}.
\end{align*}
By introducing a latent class index $s_i\in\{1,\ldots,k\}$ for the $i$th subject, the multivariate responses ${\bf x}_i=(x_{i1},\ldots,x_{ip})'$ are conditionally independent given $s_i$.  Instead of conditioning on a fixed $k$, \cite{DunsonXing09} developed a nonparametric Bayes approach that lets
\begin{align}
\bm{\pi} &= \sum_{h=1}^{\infty} \nu_{h} \Psi_h, \ \ \ \ \Psi_h = \bm{\psi}_h^{(1)} \otimes \cdots \otimes \bm{\psi}_h^{(p)}, \label{eq:dx1} \\
\bm{\psi}_h^{(j)}&\sim \text{Dirichlet}(a_{j1},\ldots,a_{jd_j}), \ \text{independently for} \ j=1,\ldots, p, \nonumber \\ 
& \hspace{46mm} h=1,\ldots,\infty, \nonumber \\
\nu_{h} &= V_h \prod_{l<h} (1 - V_l), \nonumber \\
V_h &\sim \text{beta}(1, \alpha), \ \text{independently for} \ h=1,\ldots,\infty, \nonumber 
\end{align}
where $a_{jl}>0$ for $l=1,\ldots,d_j$ and $\alpha>0$.  Although (2) allows infinitely many components, the number $k_n$ occupied by the $n$ subjects in the sample will tend to be $k_n << n$, so few components will be occupied. The model corresponds to a Dirichlet process mixture of product multinomial distributions relying on a stick-breaking representation (\cite{Sethuraman94}). A prior is induced on the joint pmf which has large support in the sense of assigning positive probability to $L_1$ neighborhoods of any true joint pmf.

\subsection{Modeling of time-indexed multivariate categorical data} \label{sq:extention}

Relying on the DX type probabilistic Parafac factorization, we propose a new nonparametric Bayes approach for time-indexed large sparse contingency tables.  In a dynamic setting, we obtain the time-indexed multivariate response ${\bf x}_{ti}=(x_{ti1},\ldots,x_{tip})'$, $x_{tij} \in \{1,\ldots,d_j\}$, for the $i$th subject at time $t$ for $i=1,\ldots, n_t$, $t=1,\ldots, T$ and $j=1,\ldots, p$.  At time $t$ we have a probability tensor $\pi_t$ for the multivariate categorical response given by
\begin{align*}
\bm{\pi}_t = \left\{ \pi_{tc_1\cdots c_p}, \, c_j=1,\ldots,d_j, \, j=1,\ldots,p \right\} \in \Pi_{d_1\cdots d_p}
\end{align*}
where $\pi_{tc_1\cdots c_p}=P(x_{ti1}=c_1,\ldots,x_{tip}=c_p)$ is a cell probability at time $t$.  Relying on the probabilistic Parafac factorization, each probability tensor $\bm{\pi}_t$ can be expressed as a mixture of product multinomials
\begin{align}
\bm{\pi}_t &= \sum_{h=1}^{k_t} \nu_{th} \Psi_{th}, \ \ \ \ \Psi_{th} = \bm{\psi}_{th}^{(1)} \otimes \cdots \otimes \bm{\psi}_{th}^{(p)} \label{eq:pit}
\end{align}
where $k_t\in \mathbb{N}$, $\bm{\nu}_t=(\nu_{t1},\ldots,\nu_{tk_t})'$ is a probability vector, $\Psi_{th}\in \Pi_{d_1\cdots d_p}$ and $\bm{\psi}_{th}^{(j)}=(\psi^{(j)}_{th1},\ldots,\psi^{(j)}_{thd_j})'$ is a $d_j\times 1$ probability vector for $h=1,\ldots,k_t$.  Letting $s_{ti} \in \{1,\ldots,k_t\}$ denote a latent class index for the $i$th subject at time $t$, the observations ${\bf x}_{ti}$ are conditionally independent given $s_{ti}$.

To borrow information across time, we place a dynamic structure on the probability tensor $\bm{\pi}_t$ in (\ref{eq:pit}) assuming time varying weights $\nu_{th}$ and static atoms $\psi_{th}^{(j)} = \psi_h^{(j)}$. Time dependence is induced in the weights through a state space model, which assumes that stick-breaking increments on $\nu_{th}$ arise through transforming Gaussian autoregressive processes using a monotone differentiable link function $g: \Re \to (0,1)$. This characterization is motivated by the probit stick-breaking process (\cite{ChungDunson09}; \cite{RodriguezDunson11}), and leads to a parsimonious but flexible characterization of time-dependence in joint pmfs underlying large, sparse contingence tables.

Similarly to expression (2), we develop a nonparametric Bayes approach that sets the number of components to $k_t = \infty$, though the number of occupied components will tend to be much less than the sample size and can vary across time.  The specific model is
\begin{align}
\bm{\pi}_t &= \sum_{h=1}^{\infty} \nu_{th} \Psi_h, \ \ \ \ \Psi_h = \bm{\psi}_h^{(1)} \otimes \cdots \otimes \bm{\psi}_h^{(p)}, 
\label{eq:new1} \\
\bm{\psi}_h^{(j)}&\sim \text{Dirichlet}(a_{j1},\ldots,a_{jd_j}), \ \text{independently for} \ j=1,\ldots, p, \label{eq:new2} \\ 
& \hspace{46mm} h=1,\ldots,\infty, \nonumber \\
\nu_{th}&=g(W_{th})\prod_{l<h}\{1-g(W_{tl})\}, \label{eq:new3} \\
W_{th} &= \alpha_{th} + \varepsilon_{th}, \ \ \varepsilon_{th} \sim N(0, \sigma_{\varepsilon}^2), \label{eq:new4}  \\
\alpha_{th} &= \mu + \phi \alpha_{t-1h} + \eta_{th}, \ \ \eta_{th} \sim N(0, \sigma^2_{\eta}), \label{eq:new5}
\end{align}
where $|\phi|<1$, $\{\varepsilon_{th}\}$ and $\{\eta_{th}\}$ are sequences of independently normally distributed random variables with mean 0 and variance $\sigma_{\varepsilon}^2$ and $\sigma_{\eta}^2$ respectively.  The parameter $\phi$ controls the autocorrelation over time in the weights $\nu_{th}$ on the different components. For sake of parsimony and simplicity in modeling and computation, we include a single time-stationary correlation parameter $\phi$ instead of allowing dependence to be time or element specific. In the limiting case in which $\phi = 0$, the weights $\nu_{th}$ will be modeled as independent. This does not mean that independent priors are placed on the unknown joint pmfs at each time, as the incorporation of common atoms automatically induces some degree of {\em a priori} dependence. However, in applications one typically expects that the joint pmfs will be quite similar over time, and by using varying weights one does not rule out arbitrarily large changes in the pmfs over time. When $\phi$ is close to one, there will be very high time dependence in the weights, leading to effective collapsing on a model that assumes a single time stationary joint pmf. For the initial state variables, we assume the stationary distributions, $\alpha_{1h} \sim N(\mu/(1-\phi), \sigma_{\eta}^2/(1-\phi^2))$ independently for $h=1,\ldots, \infty$.  Also, we choose priors  $\mu \sim N(\mu_0, \sigma^2_0)$, $\phi \sim U(-1, 1)$, $\sigma^2_{\varepsilon} \sim IG(m_{\varepsilon}/2, S_{\varepsilon}/2)$ and $\sigma^2_{\eta}\sim IG(m_{\eta}/2, S_{\eta}/2)$ respectively.

Expressions (4)-(8) induce a prior on the time-dependent joint pmfs, but it is not immediately obvious how the chosen hyperpriors in the hierarchical specification impact the properties of the prior for $\{ \bm{\pi}_t \}$. In particular, it is important to obtain characterizations of the moments of the induced prior for the cell probabilities, as well as the prior covariance between different elements and across time. Such expressions are provided in Lemma 1, with the proof provided in Appendix A. Lemma 2 shows that the prior is well defined in the sense that $\sum_{h=1}^{\infty} \nu_{th}$ converges to one almost surely.

Lemma 1. The expectation, variance and covariance of the joint prior on the elements of $\{ \bm{\pi}_t \}$ induced through (4)-(8) are
\begin{align*}
&E\{\pi_{tc_1\cdots c_p}\} = \prod_{j=1}^p \frac{a_{jc_j}}{\hat{a}_{j}}, \hspace{5mm}
V\{\pi_{tc_1\cdots c_p}\} = \left( \prod_{j=1}^p \frac{a_{jc_j}(a_{jc_j}+1)}{\hat{a}_j(\hat{a}_j+1)} - \prod_{j=1}^p \frac{a^2_{jc_j}}{\hat{a}^2_{j}} \right)\left( \frac{\beta_2 }{2\beta_1-\beta_2} \right),\\
&Cov\{\pi_{tc_1\cdots c_p}, \pi_{t+k c'_1\cdots c'_p}\} = \left( \prod_{j=1}^p \frac{a_{jc_j} \{ a_{jc'_j} + 1(c_j=c'_j) \}}{\hat{a}_j (\hat{a}_j + 1)} - \prod_{j=1}^p \frac{a_{jc_j} a_{jc'_j} }{\hat{a}^2_j} \right)\left( \frac{\gamma_k}{ 2\beta_1-\gamma_k } \right),
\end{align*}  
where $\beta_{1} = E\{g(W_{th})\}$, $\beta_{2} = E\{g^2(W_{th})\}$, $\gamma_k = E \left\{ g(W_{th}) g(W_{t+kh})\right\}$, $\hat{a}_j=\sum_{l=1}^{d_j} a_{jl}$ and $1(\cdot)$ is an indicator function.

The expectation of cell probabilities can be expressed as the product of expectations of Dirichlet priors for atoms.  The variance and covariance are expressed as the product of two terms, the first one is related to atoms and the second one comes from time varying weights.  As $\mu \rightarrow \infty$, then $\beta_2 / (2\beta_1-\beta_2) \rightarrow 1$ and $\gamma_k / (2\beta_1-\gamma_k) \rightarrow 1$, and the variance and covariance will be influenced only by atoms.  In such a case, the measure corresponding to the stick-breaking process will become a point mass at a random atom almost surely.  In addition, $\beta_1$, $\beta_2$ and $\gamma_k$ do not depend on time $t$, hence all expectation, variance and covariance are independent of $t$ though the covariance depends on the time difference $k$.  Also, the covariance between cell probabilities with $c_j=c'_j$ for all $j$ is always positive and, on the other hand, those with $c_j\neq c'_j$ for all $j$ have negative covariance.  In a special case in which the hyperparameters in the Dirichlet prior are $a_{j1}= \cdots = a_{jd_j}=a$ the variance and covariance is zero in the limit as $a \to \infty$. The proof is in Appendix A.

Lemma 2. $\sum_{h=1}^{\infty} \nu_{th} = 1$ almost surely.

Lemma 2 is important in showing that the prior is well defined. The proof is in Appendix B.

Our proposed prior setting is parsimonious but highly flexible in the sense that the induced prior assigns positive probability in arbitrarily small neighborhoods of any true data-generating pmf.  
Let $\Pi$ denote the space having elements of the form  $\bm{\pi}=\{ \bm{\pi}_t \in \Pi_{d_1\cdots d_p}, \, t\in\{1,\ldots,T\} \}$. We show in Theorem 1 that the proposed prior has large support on $\Pi$.

\vspace{3mm}
\noindent
\textit{Theorem 1.}
Let $\mathcal{Q}$ denote the prior on $\Pi$ through the proposed model and $\mathcal{N}_{\epsilon}(\bm{\pi}^0)$ denote an $L_1$ neighborhood around an arbitrary $\bm{\pi}^0 \in \Pi$.  Then for any $\bm{\pi}^0\in\Pi$ and $\epsilon > 0$, the prior assigns positive probability in the $\epsilon$-neighborhood, $\mathcal{Q}\left\{ \mathcal{N}_{\epsilon}(\bm{\pi}^0)\right\}>0$.
\vspace{3mm}

Since the proposed prior is defined on a space with finitely many components, a straightforward extension of theorem 4.3.1 in \cite{GhoshRamamoorthi03} ensures that the posterior concentrates in arbitrary small neighborhoods of any true data-generating distribution as the sample size increases.

\section{MCMC algorithm for posterior computation}

For posterior computation in DP mixtures, one common approach is marginalizing out the random probability measure with the Polya urn scheme (\cite{BushMacEachern96}).  Avoiding marginalization, \cite{IshwaranJames01} developed the blocked Gibbs sampler relying on truncation approximation of the stick-breaking representation.  Without truncation, \cite{Walker07} and \cite{PapaspiliopoulosRoberts08} proposed the slice sampler and retrospective MCMC methods respectively.  Though the slice sampler is simpler to implement, conditional constraints on sticks can cause slow mixing of the chain.  \cite{KalliGriffinWalker11} proposed a more efficient slice sampler avoiding such a mixing problem.

Relying on a slice sampler related to \cite{KalliGriffinWalker11}, we developed a simple and efficient MCMC algorithm for the proposed model.  In the motivating application, we have two types of missing data, design-based missingness and
individual-specific missingness.  We assume missing at random for both cases and handle the missing data using missingness indicators, $m_{ti} = (m_{ti1},\ldots, m_{tip})'$, with $m_{tij} = 1$ if variable $j$ is missing for subject $i$ at time $t$.  In addition, we introduce latent variables $u_{t}=(u_{t1},\ldots,u_{tn_t})'$ for the slice sampler.  The likelihood of $\{u_{t}\}$ and $\{{\bf x}_t\}$ given $\{m_{ti}\}$, $\{\bm{\nu}_{t}\}$ and $\left\{\bm{\psi}_{h}^{(j)}\right\}$ can be expressed as
\begin{align*}
\prod_{t=1}^{T} \prod_{i=1}^{n_t} \left\{ \sum_{h=1}^{\infty} 1(u_{ti} < \nu_{th}) \prod_{j:\,m_{tij}=0} \prod_{l=1}^{d_j} \left( \psi_{hl}^{(j)} \right)^{1(x_{tij}=l)} \right\}.
\end{align*}
This representation is consistent with the original model setting if latent variables $\{u_{t}\}$ are marginalized out.  In a special case in which $g$ is a probit link function, the data augmentation approach in \cite{AlbertChib01} can improve efficiency of the posterior sampling by introducing independent normal latent variables $\{z_{tih}\}$ with mean $W_{th}$ and variance 1 satisfying 
\begin{align*}
P(z_{tih} > 0, \, z_{til} \leq 0, \, l<h) &=\Phi(W_{th}) \prod_{l<h} \{ 1 - \Phi(W_{th}) \} = \nu_{th} = P(s_{ti}=h).
\end{align*}
We propose the following MCMC sampling steps:
\begin{enumerate}
\item  For $h=1,\ldots,k^*$, with $k^*=\max\{s_{ti}\}$, update $\bm{\psi}_h^{(j)}$ from the following Dirichlet full conditional posterior distribution,
\begin{align*}
\text{Dirichlet}\left( a_{j1} +  \sum_{(t,i) \in A_{jh}} 1(x_{tij}=1),\, \ldots,\, a_{jd_j} + \sum_{(t,i) \in A_{jh}} 1(x_{tij}=d_j) \right).
\end{align*}
where $A_{jh}=\{(t,i): m_{tij}=0,\, s_{ti}=h\}$.

\item  Update $z_{tih}$ from the marginal (w.r.t. $u_{ti}$) conditional posterior distribution,
\begin{align*}
z_{tih}\,|\,\cdots \sim \begin{cases}
                N_{-}(W_{th}, 1) & h<s_{ti}, \\
                N_{+}(W_{th}, 1) & h=s_{ti}, 
              \end{cases}			
\end{align*}
where $N_{-}(W_{th}, 1)$ and $N_{+}(W_{th}, 1)$ denote the normal distributions with mean $W_{th}$ and variance 1 truncated on $(-\infty, 0]$ and $(0, \infty)$ respectively.

\item  Update $W_{th}$ from the normal marginal (w.r.t. $u_{ti}$) conditional posterior distribution, $N(\hat{W}_{th},\sigma^2_{W_{th}})$ where
\begin{align*}
\hat{W}_{th}=\sigma^2_{W_{th}} \left(\sum_{i:s_{ti} \geq h}^{n_t}z_{tih}+\sigma^{-2}_{\varepsilon}\alpha_{th}\right), \ \ 
\sigma^2_{W_{th}}= \frac{1}{\sum_{i=1}^{n_t} 1(s_{ti} \geq h) + \sigma^{-2}_{\varepsilon}}. 
\end{align*} 

\item  Update $u_{ti}$ from the full conditional distribution, Uniform$(0, \nu_{ts_{ti}})$.

\item  Update $s_{ti}$ from the multinomial full conditional distribution,
\begin{align*}
Pr(s_{ti}=h\,|\, \cdots) = \frac{1(h\in B_{ti}) \prod_{j:m_{tij}=0} \psi_{hx_{tij}}^{(j)} }{ \sum_{l\in B_{ti}} \prod_{j:m_{tij}=0} \psi_{lx_{tij}}^{(j)} },
\end{align*}
where $B_{ti}=\{h:\, \nu_{th} > u_{ti}\}$.  To identify the elements in $\{B_{ti}\}$, we first update $\alpha_{th}$ and $W_{th}$ for $t=1,\ldots,T$ and $h=1,\ldots,\tilde{k}$ where $\tilde{k}$ is the smallest number with $\sum_{h=1}^{\tilde{k}}\nu_{th}>1-\min\{s_{ti}\}$ for all $t$.

\item  For $h=1,\ldots,k^*$, update $\alpha_{th}$ using the forward filtering backward sampling algorithm by \cite{FruhwirthSchnatter94} and \cite{CarterKohn94}, or Kalman filter and the simulation smoother by \cite{deJongShephard95} and \cite{DurbinKoopman02}. 

\item  Update $\mu$ from the conditional posterior, $N(\mu_*, \sigma^2_{\mu})$ where $\mu_*=\sigma^2_{\mu}(\hat{\sigma}^{-2}\hat{\mu}+\sigma_{0}^{-2}\mu_{0})$, $\sigma^2_{\mu}=(\hat{\sigma}^{-2}+\sigma_{0}^{-2})^{-1}$ and 
\begin{align*}
\hat{\mu} = \frac{\sum_{h=1}^{k^*} \sum_{t=2}^T ( \alpha_{th}- \phi\alpha_{t-1h} ) + (1+\phi) \sum_{h=1}^{k^*} \alpha_{1h} }{k^*\left\{T-1+(1+\phi)/(1-\phi)\right\}}, \ \ 
\hat{\sigma}^2= \frac{\sigma^2_{\eta} }{k^*\left\{T-1+(1+\phi)/(1-\phi)\right\}}.
\end{align*}

\item  Update $\phi$ using the independence MH algorithm in which the proposal distribution is constructed relying on the mode and Hessian of the logarithm of the conditional posterior densities $\pi(\phi|\cdots)$.  First, we compute $\hat{\phi}$ which maximizes (or approximately maximizes) the conditional posterior density. Then, we generated a candidate from a truncated normal distribution $TN_{(-1,1)}(\phi_*,\sigma_{\phi}^2)$, where
	\begin{eqnarray*}
	\phi_*=\hat{\phi}+\sigma_{\phi}^{2}\left.\frac{\partial \log\pi(\phi|\cdots)}{\partial\phi}\right|_{\phi=\hat{\phi}},\hspace{1em}
	\sigma_{\phi}^{2}=\left\{ -\left.\frac{\partial \log\pi(\phi|\cdots)}{\partial^2\phi}\right|_{\phi=\hat{\phi}} \right\}^{-1}. \label{eq:phi}
	\end{eqnarray*}

\item  Update $\sigma^2_{\varepsilon}$ from the conditional distribution, $IG(\hat{m}_{\varepsilon}/2, \hat{S}_{\varepsilon}/2)$ where $\hat{m}_{\varepsilon}=Tk^*+m_{\varepsilon}$ and $\hat{S}_{\varepsilon}=\sum_{t=1}^T\sum_{h=1}^{k^*}( W_{th} - \alpha_{th} )^2 + S_{\varepsilon}$.

\item  Update $\sigma^2_{\eta}$ from the conditional distribution, $IG(\hat{m}_{\eta}/2, \hat{S}_{\eta}/2)$ where $\hat{m}_{\eta}=Tk^*+m_{\eta}$ and $\hat{S}_{\eta}=\sum_{h=1}^{k^*} \sum_{t=2}^T( \alpha_{th} - \mu - \phi\alpha_{t-1h} )^2 + (1-\phi^2) \sum_{h=1}^{k^*} \{ \alpha_{1h} - \mu/(1-\phi) \}^2 + S_{\eta}$.
\end{enumerate}

In a case in which $g$ is another link function, we update $W_{th}$ using the independent MH algorithm, instead of step 2 and 3 above.  We generate a candidate from a normal distribution relying on the mode and Hessian of the logarithm of the conditional posterior densities of $W_{th}$.

\section{Simulation study}

In this section, we assess the impact of borrowing of information over time by comparing our proposed method to static approaches, such as Dunson and Xing (DX) (2009), applied separately at each time on simulated data.  First, we simulate time-indexed contingency tables from the model shown in expressions (\ref{eq:new1})-(\ref{eq:new5}) with $T=10$, $P=20$, $d_j=4$ for all $j$, $\mu=0$, $\phi =0.8$, $\sigma_{\varepsilon}=0.1$ and $\sigma_{\eta}=0.8$.  At the respective time points we generated 120, 110, 150, 80, 100, 120, 100, 140, 110 and 150 observations, tiny sample sizes compared with the number of cells.  For prior distributions, we assumed $\bm{\psi}_h^{(j)}\sim\text{Dirichlet}(1,\ldots,1)$, $\mu \sim N(0, 1)$, $\phi \sim U(-1, 1)$, $\sigma^2_{\varepsilon} \sim IG(2.5, 0.025)$, $\sigma^2_{\eta}\sim IG(2.5, 0.025)$.  We draw 60,000 MCMC samples after the initial 20,000 samples are discarded as a burn-in period and every fifth sample is saved.  We observed that the sample paths were stable and the sample autocorrelations dropped smoothly.  Therefore, the chains apparently converged and mixed rapidly.

\begin{figure}[H]
\centering
\includegraphics[height=15cm, width=14cm]{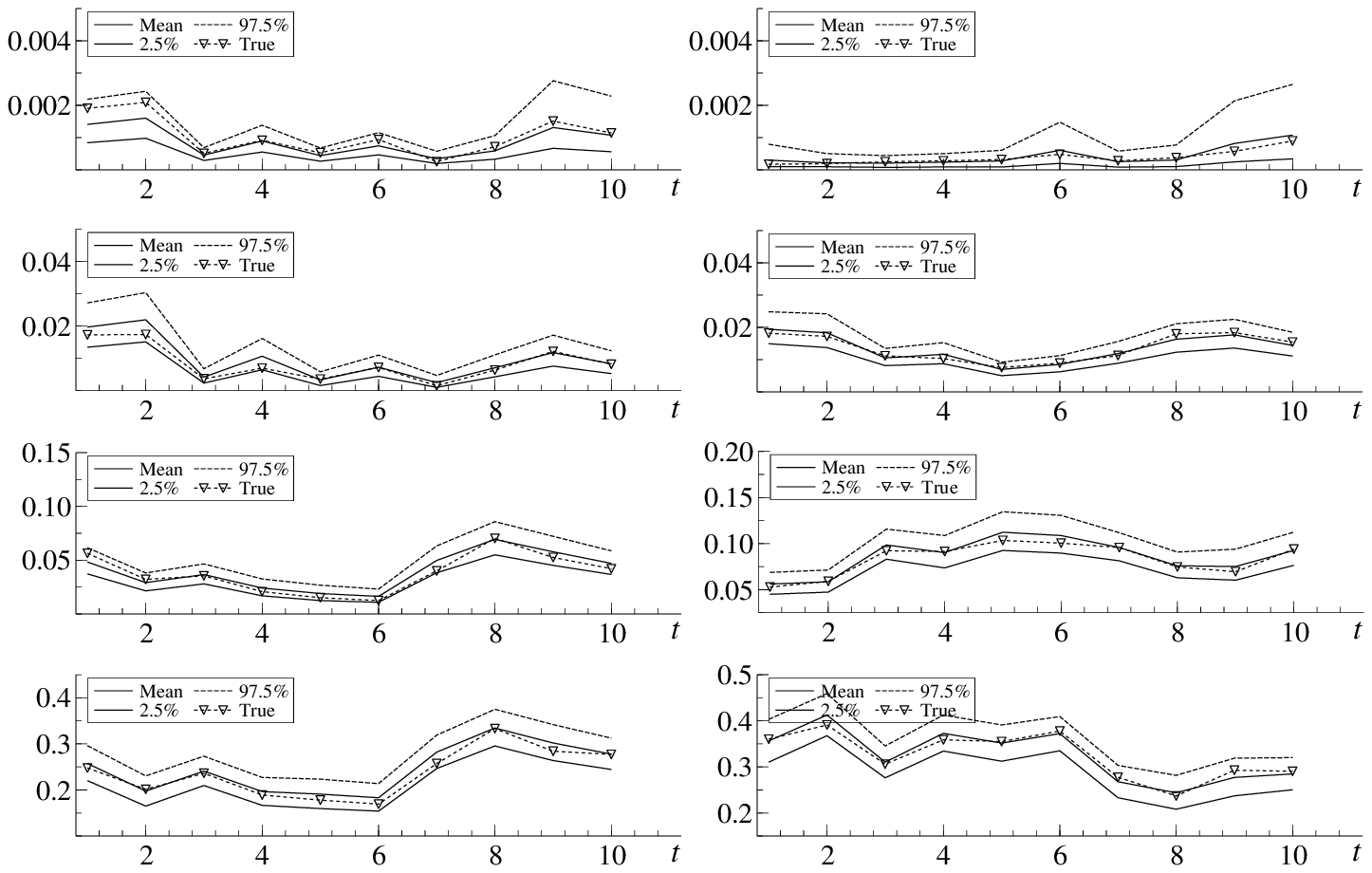}
\begin{minipage}{12cm}
{\footnotesize The first row: $P(x_{ti4}=0,x_{ti6}=1,x_{ti10}=2,x_{ti15}=3)$ and $P(x_{ti7}=2,x_{ti9}=0,x_{ti13}=3,x_{ti19}=1)$. \\
The second row: $P(x_{ti1}=2,x_{ti7}=1,x_{ti20}=3)$ and $P(x_{ti3}=3,x_{ti12}=1,x_{ti18}=0)$. \\
The third row: $P(x_{ti11}=1,x_{ti17}=1)$ and $P(x_{ti5}=2,x_{ti19}=1)$. \\
The forth row: $P(x_{ti8}=0)$ and $P(x_{ti20}=3)$.}
\end{minipage}

\caption{Estimation results of cell probabilities by the proposed method.}
\label{fig:dif1}
\end{figure}

\begin{figure}[H]
\centering
\includegraphics[height=15cm, width=14cm]{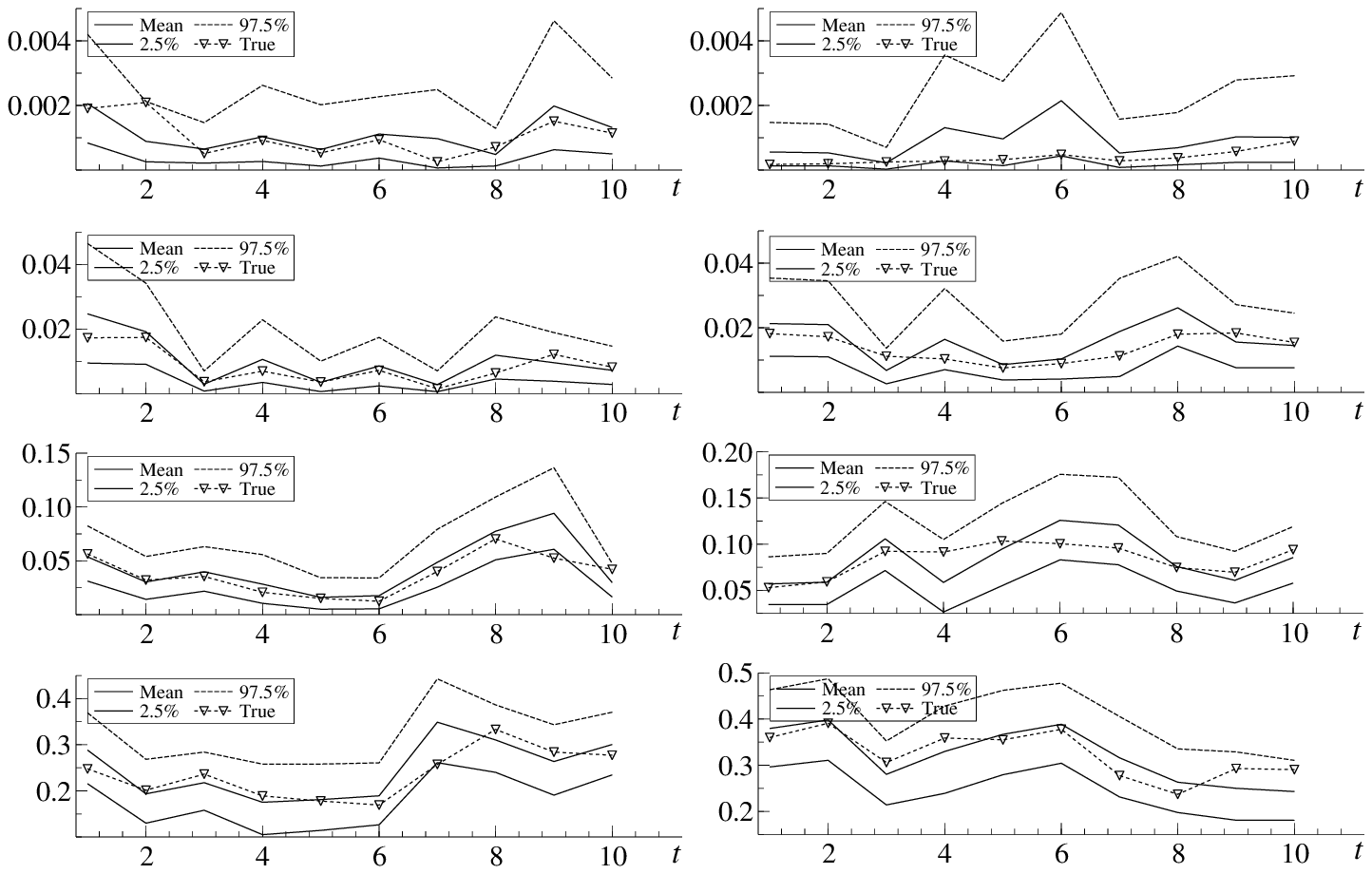}
\begin{minipage}{12cm}
{\footnotesize The first row: $P(x_{ti4}=0,x_{ti6}=1,x_{ti10}=2,x_{ti15}=3)$ and $P(x_{ti7}=2,x_{ti9}=0,x_{ti13}=3,x_{ti19}=1)$. \\
The second row: $P(x_{ti1}=2,x_{ti7}=1,x_{ti20}=3)$ and $P(x_{ti3}=3,x_{ti12}=1,x_{ti18}=0)$. \\
The third row: $P(x_{ti11}=1,x_{ti17}=1)$ and $P(x_{ti5}=2,x_{ti19}=1)$. \\
The forth row: $P(x_{ti8}=0)$ and $P(x_{ti20}=3)$.}
\end{minipage}
\caption{Estimation results of cell probabilities by DX method.}
\label{fig:dif2}
\end{figure}

We first assess performance in estimation of cell probabilities.  We picked several cells randomly and report true values, posterior means and 95\% credible intervals in Figure \ref{fig:dif1} (the proposed method) and Figure \ref{fig:dif2} (DX method).  The proposed approach covers all true values in 95\% intervals and interval widths are much narrower than for the DX approach consistently across time.

\begin{figure}[H]
  \begin{center}
    \includegraphics[height=19cm, width=15.5cm]{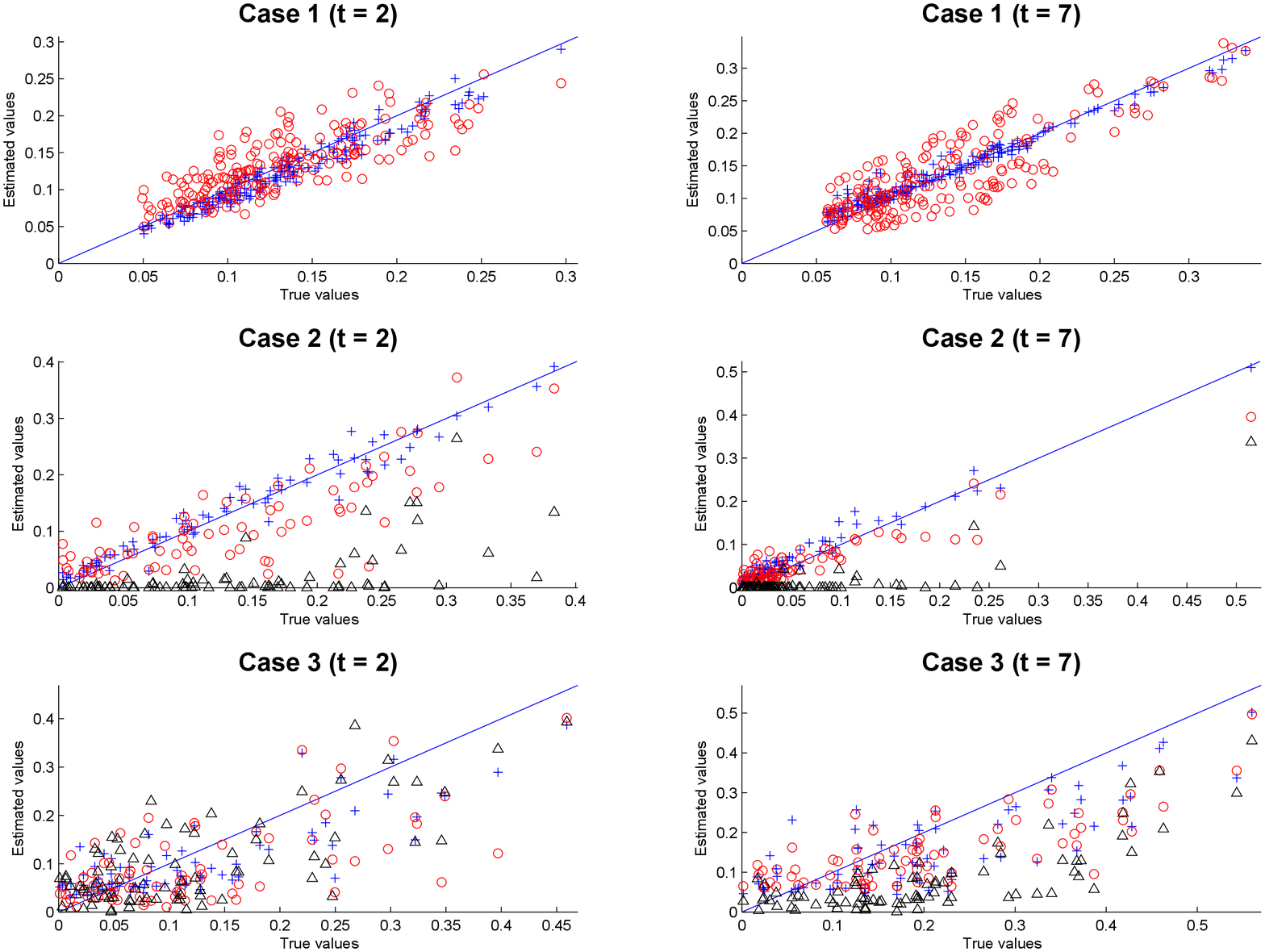}
  \end{center}
\begin{minipage}{15cm}
{\footnotesize $y$ axis represents estimated values and $x$ axis true values. Cross-shaped dots represent the proposed method, circles DX method and triangles DH method. The first, second and third rows show the results at time $t=2$ and $7$ using the first (case 1), second (case 2), third (case 3) simulation data sets.}
\end{minipage}
\caption{Plots of true and estimated values of $\rho_{tjj'}$ using the simulation data.}
\label{fig:dif3}
\end{figure}

We additionally investigate performance in estimating associations among the categorical variables using the following measure of dependence from Dunson and Xing (2009)
\begin{align}
\rho_{tjj'}^2 = \frac{1}{\min\{d_j,d_{j'}\}-1} \sum_{c_j=1}^{d_j} \sum_{c_{j'}=1}^{d_{j'}} \frac{ \left( \pi_{tc_j c_{j'}}-\bar{\psi}_{tc_j}^{(j)} \bar{\psi}_{tc_{j'}}^{(j')} \right)^2 }{\bar{\psi}_{tc_j}^{(j)} \bar{\psi}_{tc_{j'}}^{(j')}},
\label{eq:measure}
\end{align}
where $\bar{\psi}_{tl}^{(j)}\equiv P(x_{tij}=l)\approx \sum_{h=1}^{k^*}\nu_{th} \psi^{(j)}_{hl}$.  The first row of  Figure \ref{fig:dif3} reports plots of all pairs of true values ($y$-axis) and posterior means ($x$-axis) of $\rho_{tjj'}$ at time $t=2$ and $7$.  At each time point, coordinate points by our approach locate closely to the $y=x$ line, compared to widely scattered points by the DX method.  In addition, Table \ref{tb:correlation} shows correlations between true values and posterior means of $\rho_{tjj'}$.  Although correlations by the DX method are high, the proposed method consistently produces higher correlations.

\begin{table}[H]
\centering
\small
\begin{tabular}{clrrrrrrrrrr}

\hline
		& \multicolumn{1}{c}{t=1}	&	\multicolumn{1}{c}{t=2}	&\multicolumn{1}{c}{t=3}	&\multicolumn{1}{c}{t=4}	&\multicolumn{1}{c}{t=5} &\multicolumn{1}{c}{t=6} &\multicolumn{1}{c}{t=7} &\multicolumn{1}{c}{t=8} &\multicolumn{1}{c}{t=9} &\multicolumn{1}{c}{t=10} &\multicolumn{1}{c}{Total}	\\
\hline 
	Proposed 	&	0.948	&	0.977 &	0.990 & 0.977 & 0.983 & 0.986 & 0.985 & 0.965 & 0.969 & 0.968 & 0.974 \\
	DX			&	0.837	&	0.794 &	0.880 & 0.761 & 0.766 & 0.921 & 0.846 & 0.817 & 0.831 & 0.793 & 0.841 \\
\hline

\end{tabular}
\caption{Correlations between true values and posterior means of $\rho_{tjj'}$ using the first simulation data.}
\normalsize
\label{tb:correlation}
\end{table}

Log linear models provide a standard choice for the analysis of contingency tables.  However, one issue is that flexible log-linear models that accommodate arbitrary interactions among the variables and allow time dependence cannot be applied directly to large, sparse tables.  Certainly, maximum likelihood estimates typically do not exist and Bayesian methods that allow an unknown dependence structure do not scale beyond small tables.  \cite{DahindenKalischBuehlmann10} proposed an approach for high-dimensional log-linear models with interactions, which relies on solving several low-dimensional subproblems that are then combined.  An earlier approach by \cite{DahindenParmigianiEmerickBuehlmann07} instead relied on L1 penalized log-linear models allowing sparsity of tables.  Also, \cite{DahindenParmigianiEmerickBuehlmann07} proposed an efficient estimation algorithm for model selection for two level categorical variables.

As a second alternative to our proposed approach, we implemented the method of Dahinden et al. (DH) (2007) in a second simulation example with $T=8$, $P=13$ and $d_j=2$ for all $j$.  Other settings are the same as in the first simulation case.  As DH did not consider time-indexed contingency tables, we applied their approach separately at each time point using the logilasso R package, with 5-way cross validation used to choose penalty parameters.  The second row of Figure \ref{fig:dif3} and Table 2 summarize the resulting dependence measures $\rho_{tjj'}$ at time $t=2$ and $7$ for each method.  For the proposed method, the posterior means are close to true values and correlations between estimates and true values are uniformly high.  The DH method has a tendency to underestimate dependence, particularly when true values are low, and has the lowest correlation between the estimates and truth.

\begin{table}[H]
\centering
\small
\begin{tabular}{clrrrrrrrrrr}

\hline
		& \multicolumn{1}{c}{t=1}	&	\multicolumn{1}{c}{t=2}	&\multicolumn{1}{c}{t=3}	&\multicolumn{1}{c}{t=4}	&\multicolumn{1}{c}{t=5} &\multicolumn{1}{c}{t=6} &\multicolumn{1}{c}{t=7} &\multicolumn{1}{c}{t=8} &\multicolumn{1}{c}{Total}	\\
\hline 
	Proposed 	&	0.951	&	0.978 &	0.979 & 0.984 & 0.986 & 0.969 & 0.981 & 0.944 & 0.965 \\
	DX			&	0.872	&	0.803 &	0.838 & 0.599 & 0.807 & 0.884 & 0.932 & 0.827 & 0.696 \\
	DH			&	0.705	&	0.557 &	0.733 & 0.466 & 0.725 & 0.506 & 0.763 & 0.487 & 0.562 \\
\hline

\end{tabular}
\caption{Correlations between true values and posterior means of $\rho_{tjj'}$ using the second simulation data.}
\normalsize
\label{tb:correlation2}
\end{table}

Finally, to gauge robustness we also simulated data from a time-dependent log-linear model in which all the coefficients of the main effects and interactions between two variables independently follow random walk processes with variance 1 and other higher interactions are zero.  The third row of Figure \ref{fig:dif3} and Table \ref{tb:correlation3} report the estimation results.  Although we find less difference among them in this case, the proposed method still shows the best performance.

\begin{table}[H]
\centering
\small
\begin{tabular}{clrrrrrrrrrr}

\hline
		& \multicolumn{1}{c}{t=1}	&	\multicolumn{1}{c}{t=2}	&\multicolumn{1}{c}{t=3}	&\multicolumn{1}{c}{t=4}	&\multicolumn{1}{c}{t=5} &\multicolumn{1}{c}{t=6} &\multicolumn{1}{c}{t=7} &\multicolumn{1}{c}{t=8} &\multicolumn{1}{c}{Total}	\\
\hline 
	Proposed 	&	0.725	&	0.827 &	0.768 & 0.798 & 0.818 & 0.916 & 0.791 & 0.807 & 0.817 \\
	DX			&	0.642	&	0.640 &	0.726 & 0.664 & 0.611 & 0.864 & 0.769 & 0.713 & 0.724 \\
	DH			&	0.371	&	0.716 &	0.821 & 0.491 & 0.611 & 0.877 & 0.764 & 0.715 & 0.624 \\
\hline

\end{tabular}
\caption{Correlations between true values and posterior means of $\rho_{tjj'}$ using the third simulation data.}
\normalsize
\label{tb:correlation3}
\end{table}

\section{Analysis of social survey data}

In this section, we apply the proposed method to data from the General Social Survey (GSS, http://www3.norc.org/GSS+Website).  Our focus is on studying associations among demographic and preference variables over time.  We select $p=29$ categorical variables from 1994 to 2010, including gender, ethnicity, preference for particular policies and many more listed in the supplemental materials.  The GSS was conducted every two years across this time period.  The numbers of observations are 2,992 (1994), 2,904 (1996), 2,832 (1998), 2,817 (2000), 2,765 (2002), 2,812 (2004), 4,510 (2006), 2,023 (2008) and 2,044 (2010) respectively.  There are abundant missing data in which only a subset of the variables were recorded for an individual, and compared to the number of cells, the sample size is quite small at each time point.

We first compared our proposed approach to log-linear models.  Unfortunately, current methodology for fitting log-linear models that allow flexible dependence structures cannot accommodate these data due to the large sparse structure, time variation and abundant missing data.  Hence, in order to provide a comparison, we initially focused on a bivariate subset of the data consisting of religious preference ($i=1,\ldots,5$) and attitude towards abortion ($j=1,2$) from 1994 to 2010.  We consider the following log-linear Poisson models.
\begin{align*}
\text{Model 1:} \hspace{10mm} N_{tij} &\sim \text{Poisson}(N_{t}\, \mu_{ij}), 
\hspace{5mm} \log \mu_{ij} = \lambda + \lambda^{R}_{i} + \lambda^{A}_{j} + \lambda^{RA}_{ij}, 
\end{align*}
where $N_{tij}$ is count of the cell $ij$ at time $t$, $N_{t} = \sum_{i}\sum_{j} N_{tij}$, $\lambda^{R}_{i}$ is an effect of the first variable (religious preference), $\lambda^{A}_{j}$ is an effect of the second variable (view of abortion) and $\lambda^{RA}_{ij}$ is an association term.  For identifiability, we assume constraints $\lambda^{R}_{5}=\lambda^{A}_{2}=\lambda^{RA}_{5j}=\lambda^{RA}_{i2}=0$.  Model 1 assumes no time-dependence in cell probabilities $\mu_{ij}/\sum_{i'}\sum_{j'} \mu_{i'j'}$.
\begin{align*}
\text{Model 2:} \hspace{10mm} N_{tij} &\sim \text{Poisson}(N_{t}\, \mu_{tij}),
\hspace{5mm} \log \mu_{tij} = \lambda_t + \lambda^{R}_{ti} + \lambda^{A}_{tj} + \lambda^{RA}_{tij}, \\
\bm{\beta}_t &= (\lambda_t, \lambda^{R}_{t1},\ldots,\lambda^{R}_{t4},\lambda^{A}_{t1},\lambda^{RA}_{t11},\ldots,\lambda^{RA}_{t41})', \\
\beta_{tl} &= \mu_l + \phi_l \beta_{t-1l} + \varepsilon_{tl}, \ \ \varepsilon_{tl} \sim N(0, \sigma^2_l), \ \ \text{independently for} \ l=1,\ldots,10,
\end{align*}
where $\lambda^{R}_{ti}$, $\lambda^{A}_{tj}$ and $\lambda^{RA}_{tij}$ are effects of the first variable, the second variable and interactions at time $t$ respectively.  We assume $\lambda^{R}_{t5}=\lambda^{A}_{t2}=\lambda^{RA}_{t5j}=\lambda^{RA}_{ti2}=0$ at each time point and $\bm{\beta}_0=\bm{0}$ for the initial values.  Model 2 is a time dependent hierarchical model where all parameters in the log-linear model follow AR(1) process independently.

We firstly estimate all models using the data from 1994 to 2008. Then, relying on the estimated parameters, we predict the contingency table in 2010 (Table \ref{tb:contingency}).  For the proposed model, we used the same MCMC settings as in the simulation study.  For log-linear models, we estimated parameters using an MCMC algorithm where missing values are imputed from conditional probabilities given observed data at each iteration.  For example, we generate the religious preference $i$ given the view of abortion $j$ with probability $\mu_{tij}/\sum_{i'} \mu_{ti'j}$.  For priors, we assumed $\bm{\beta} =  (\lambda, \lambda^{R}_{1},\ldots,\lambda^{R}_{4},\lambda^{A}_{1},\lambda^{RA}_{11},\ldots,\lambda^{RA}_{41})'\sim N(\bm{0},I)$ for Model 1, $\mu_l\sim N(0,1)$, $\phi_l\sim U(-1,1)$ and $\sigma_l^2\sim IG(2.5,0.025)$ for all $l$ for Model 2.  Using Gibbs sampling, we generated posterior samples of $\mu_l$ and $\sigma^2_j$ from normal and Inverse-Gamma distributions respectively.  For $\bm{\beta}$, $\phi_l$, $\bm{\beta}_t$, we used a MH algorithm in which candidates were generated from normal distributions relying on  the mode and Hessian of the logarithm of the conditional posterior densities.  We generated 10,000 MCMC samples after the 1,000 burn-in for Model 1 and 20,000 MCMC samples after the 2,000 burn-in for Model 2 and, for both cases, every fifth sample was saved.

We generated replications at every fifth MCMC iteration and computed average of the following predictive criteria,
\begin{align*}
\text{Absolute deviation (AD):}&\hspace{5mm}  \sum_{i=1}^5 \sum_{j=1}^2 \left| N^{rep}_{ij} - N^{obs}_{ij} \right|, \\
\text{Mean absolute percentage error (MAPE):}&\hspace{5mm} \frac{1}{10} \sum_{i=1}^5 \sum_{j=1}^2 \left| \frac{N^{rep}_{ij} - N^{obs}_{ij}}{N^{obs}_{ij}} \right|,
\end{align*}
where $N^{rep}_{ij}$ and $N^{obs}_{ij}$ are the replication and observation of count of the cell $ij$ respectively.  To keep the same total number of replications among all methods, predictions are generated from cell probabilities $\mu_{ij}/\sum_{i'}\sum_{j'} \mu_{i'j'}$ for Model 1 and $\mu_{2010ij}/\sum_{i'}\sum_{j'} \mu_{2010i'j'}$ for Model 2.  Table \ref{tb:pred-result} reports the prediction results.  Although Model 2 produces better performance than Model 1 by incorporating time-dependence, the proposed method clearly outperforms log-linear models in terms of both predictive criteria.

\begin{table}[H]
\centering
\small
\begin{tabular}{crrrrrr}
\hline
		& \multicolumn{1}{c}{Protestant}	&	\multicolumn{1}{c}{Catholic}	&\multicolumn{1}{c}{Jewish}	&\multicolumn{1}{c}{None}	&\multicolumn{1}{c}{Other} &\multicolumn{1}{c}{Total}	\\
\hline 
	Agree		 	&	216	&	103 &	21 & 137 & 60 & 537   \\
	Disagree		&	372	&	182 &	7  & 81  & 47 & 689   \\
	Total			&	588	&	285 &	28 & 218 & 107 & 1226   \\
\hline
\end{tabular}
\caption{Contingency table of the religious preference and view of abortion in 2010.}
\normalsize
\label{tb:contingency}
\end{table}

\begin{table}[H]
\centering
\small
\begin{tabular}{crrrrr}

\hline
		& \multicolumn{1}{c}{Proposed}	&	\multicolumn{1}{c}{Model 1}	&\multicolumn{1}{c}{Model 2}	\\
\hline 
	AD		 	&	194.4	&	208.6  	&	204.5   	   \\
	MAPE		&	0.216	&	0.232	&	0.227   	   \\
\hline

\end{tabular}
\normalsize
\caption{Prediction results.}
\label{tb:pred-result}
\end{table}

Next, we apply the proposed method to all 29 categorical variables.  We generated 30,000 MCMC samples after the initial 10,000 samples are discarded as the burn-in and every fifth sample are saved.  We observed the sample paths are stable and the sample autocorrelations are small.  Table \ref{tb:real-para} shows the estimation result of parameters in the time dependent stick-breaking processes.  Concerning the measure of time dependence $\phi$, the posterior mean is close to 1 and the 95\% credible interval locates near 1, which means the weights of the stick-breaking processes have strong time dependence over time.

\begin{table}[H]
\centering
\small
\begin{tabular}{clrr}

\hline
	Parameter	& \multicolumn{1}{c}{Mean}	&	\multicolumn{1}{c}{Stdev.}	&\multicolumn{1}{c}{95\% interval}	\\
\hline 
	$\mu$	&	-0.012	&	0.004 &	[-0.023, -0.005]	\hspace{2mm} \\
	$\phi$	&	0.988	&	0.004 &	[0.978, 0.994]	\hspace{2mm} \\
	$\sigma_{\varepsilon}$	&	0.062	&	0.009 &	[0.046, 0.082]	\hspace{2mm} \\
	$\sigma_{\eta}$	&	0.126	&	0.011 &	[0.104, 0.149] \hspace{2mm} \\
\hline

\end{tabular}
\caption{Estimation result of parameters in the proposed stick-breaking process.}
\normalsize
\label{tb:real-para}
\end{table}

\begin{figure}[H]
  \begin{center}
    \subfigure{\includegraphics[scale=0.7]{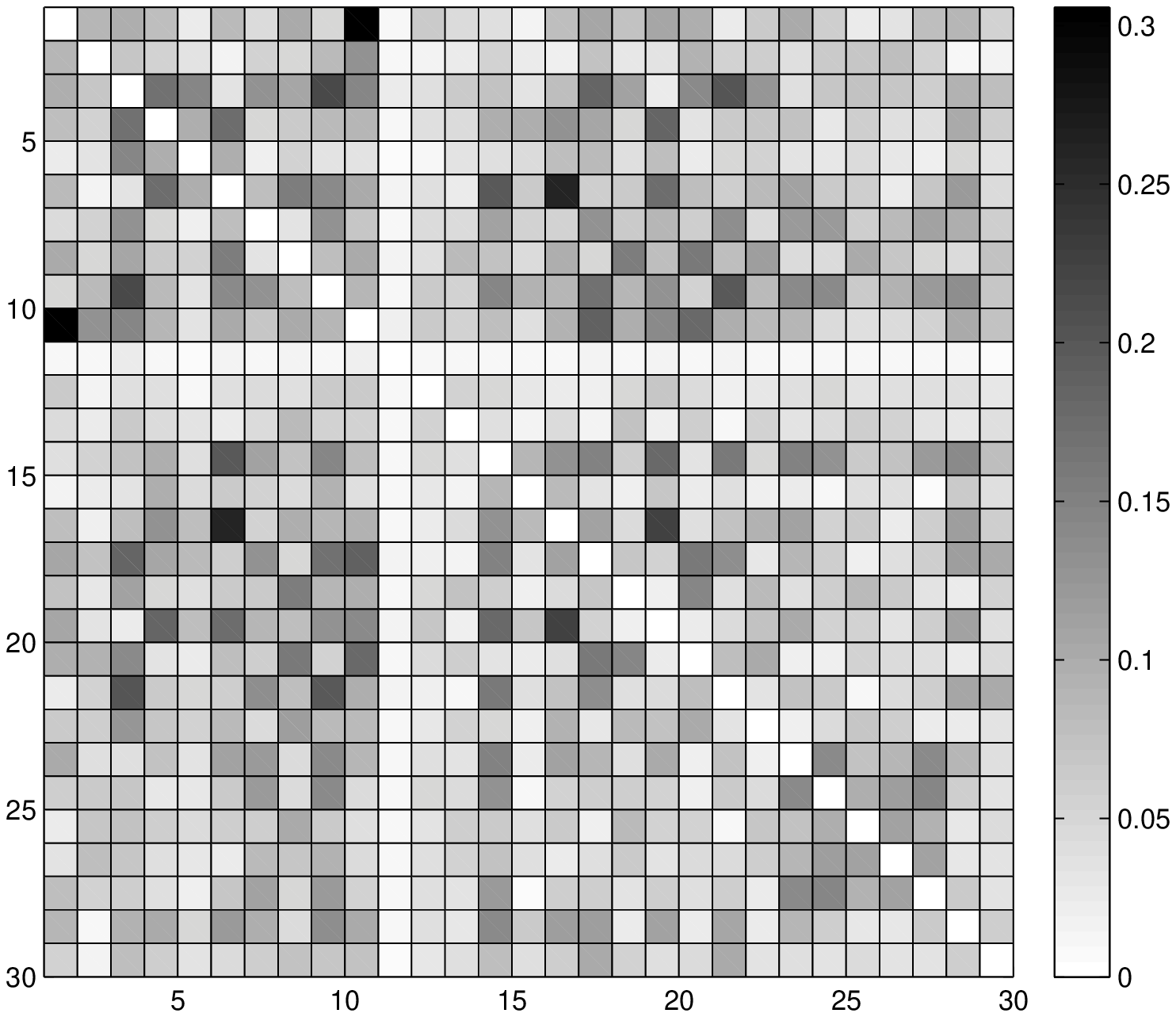}}
	\subfigure{\includegraphics[scale=0.7]{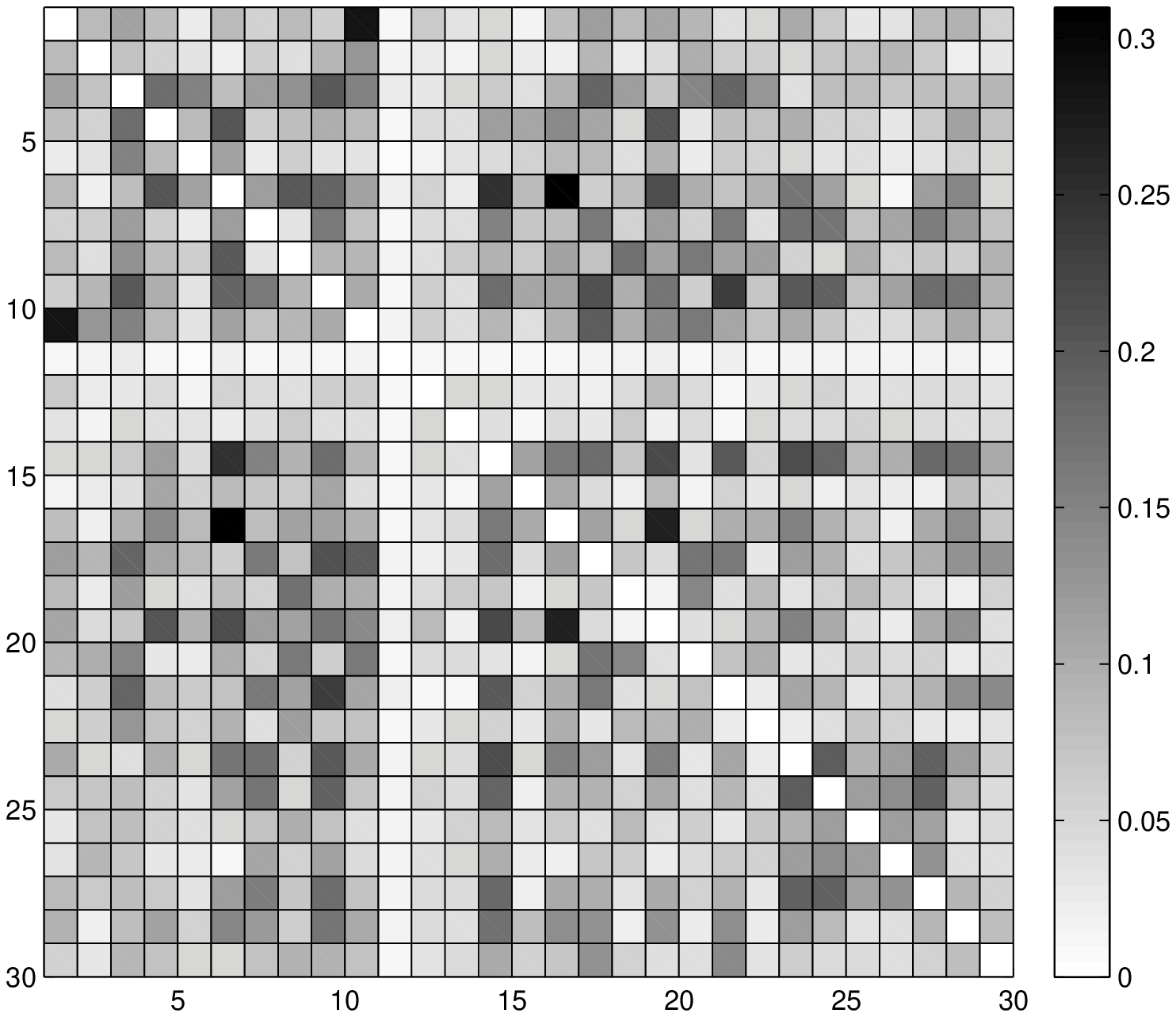}}
  \end{center}
\caption{Posterior means of $\rho_{tjj'}$ in 2002 (above) and 2010 (below).}
\label{fig:table}
\end{figure}

Then, we investigate cross interactions among the variables over time.  Figure \ref{fig:table} show the posterior means of $\rho_{tjj'}$ for all pairs in 2002 and 2010.  Additional results for other years are included in the supplemental materials.  We find the structure of interactions is complex at each time point. Also, though each interaction gradually changes over time, all tables look similar to one another, implying they have close dependence.  This is consistent with the result of the strong dependent weights in the stick-breaking processes.  Some categorical variables such as Race [$j=3$], Attitude toward abortion [6], Political party affiliation [9] and Think of self as liberal or conservative [14] intricately correlate with many other variables.  On the other hand, zodiac [11] shows little interactions with all other variables.  Among all pairs of variables, \{Age [1], Marital status [10]\}, \{Attitude toward abortion [6], Attitude toward homosexual [16]\} and \{Attitude toward homosexual [16], Attitude toward Marijuana [19]\} show strong interactions in the whole period.  Also, we observed several pairs of variables showing relatively close interactions over time, such as \{Attitude toward abortion [6], Think of self as liberal or conservative [14]\}, \{Race [3], Political party affiliation [9]\} and \{Marital status [10], Having gun [17]\}.  In addition, the views of government expense show moderate interactions, especially to the environment [23], nation's health [24], halting the rising crime [25], dealing with drug addiction [26] and education system [27].

\begin{figure}[H]
\centering
\includegraphics[height=14cm, width=16cm]{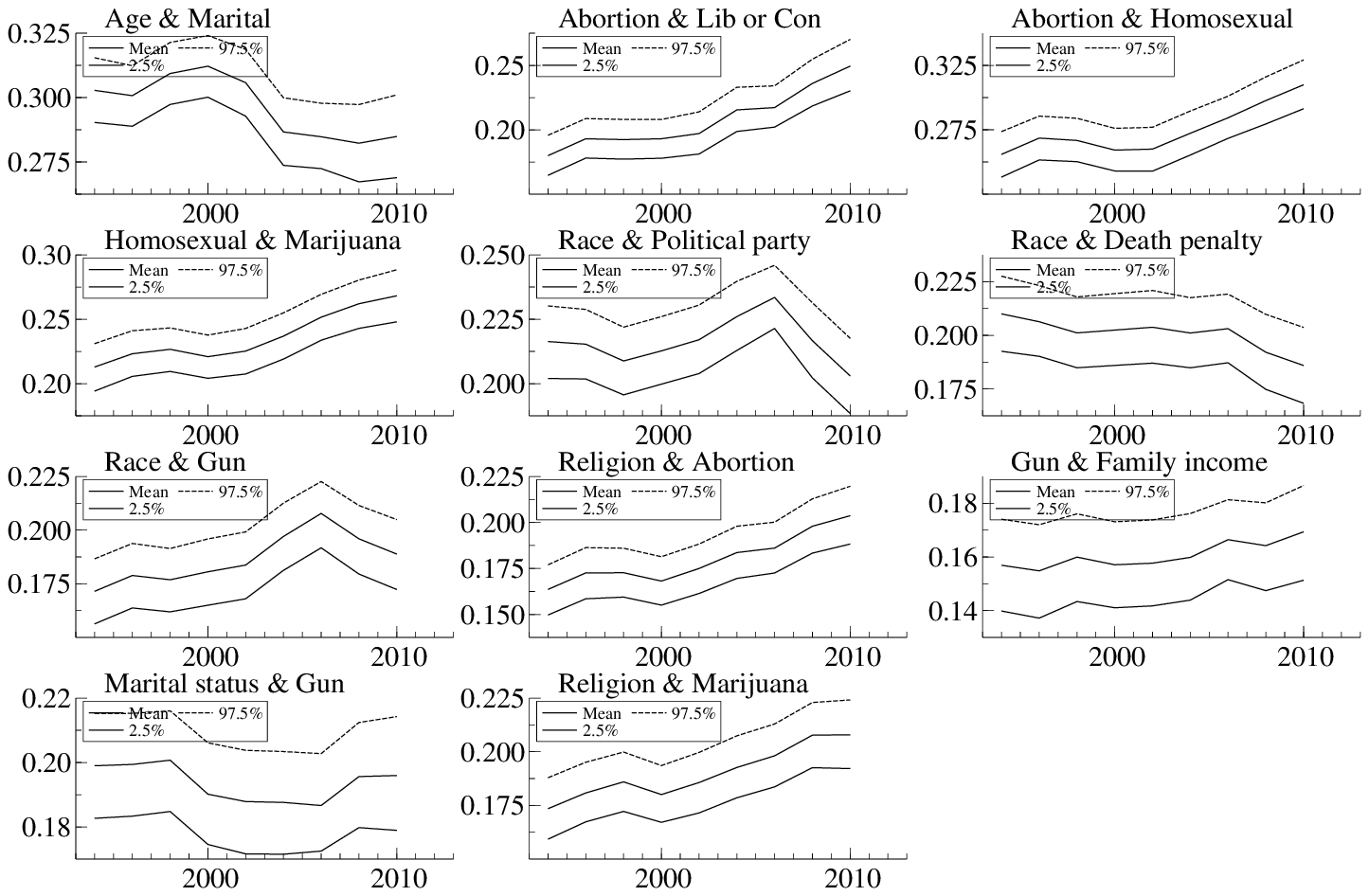}
\begin{minipage}{16cm}
{\footnotesize The first row: (Age group, Current marital status), (Attitude toward abourtion, Think of self as liberal or conservative) and (Attitude toward abourtion, Attitude toward homosexual sex relations). \\ 
The second row: (Attitude toward homosexual sex relations, Should Marijuana be made legal), (Race, Political party affiliation) and (Race, Favor or oppose death penalty for murder). \\ 
The third row: (Race, Have gun in home), (Religious preference, Attitude toward abourtion) and (Have gun in home, Total family income). \\ 
The fourth row: (Current marital status, Have gun in home) and (Religious preference, Should Marijuana be made legal).}
\end{minipage}
\caption{Estimation results of $\rho_{tjj'}$ for several pairs.}
\label{fig:rhoplot1}
\end{figure}

Next, we study trends of dependence between categorical variables.  Figure \ref{fig:rhoplot1} reports the posterior means and 95\% credible intervals of $\rho_{tjj'}$ for pairs with close interactions.  We observed various patterns of time paths.  For \{Age, Marital status\}, the interaction increased around 2000 then declined sharply to a lower level.  \{Race, Political party affiliation\} and \{Race, Having gun\} have peaks in 2006 and the interactions have steeply decreased after that.  In addition, we can see similar trends in \{Attitude toward abortion, Think of self as lib or con\}, \{Attitude toward abortion, Attitude toward homosexual\}, \{Attitude toward homosexual, Attitude toward Marijuana\}, \{Religion, Attitude toward abortion\} and \{Religion, Attitude toward Marijuana\}.  The interactions have roughly increased over time, especially in the 2000s.  On the other hand, the dependence in \{Race, Death penalty for murder\} decreased at first and kept stable in the middle of the period then declined again.  \{Having gun, Family income\} gradually increased over the period but the difference is small.  For \{Marital status, Having gun\}, the interaction dropped in the middle of the period but recovered recently at the same level as the beginning.

\section{Discussion}

We have demonstrated that the proposed approach is useful in analyzing time-indexed large sparse contingency tables. One interesting extension is to accommodate joint modeling of mixed scale variables consisting of not only categorical data but also continuous and count variables. In such a case, one can potentially model the observed data vector for the $i$th subject at time $t$, $y_{ti} = (y_{ti1},\ldots,y_{tip})'$, as conditionally independent given latent class variables $x_{ti} = (x_{ti1},\ldots,x_{tip})'$, with $x_{ti}$ modeled exactly as proposed in this article. For example, consider the simple case in which $p=2$ with $y_{ti1} \in \Re$ continuous and $y_{ti2} \in \{1,\ldots,d_2\}$ categorical. Then, one can let $y_{ti1} \sim N( \mu_{x_{ti1}}, \sigma_{x_{ti1}}^2 )$ and $y_{ti2}=x_{ti2}$, with the proposed probabilistic tensor factorization approach flexibly accommodating dependence in $y_{ti1}$ and $y_{ti2}$ through dependence in $x_{ti1}$ and $x_{ti2}$. The induced marginal distribution for the continuous variable $y_{ti1}$ will be a mixture of normals, with the probability weight on each component potentially varying with the categorical variable $y_{ti2}$. This same strategy can be generalized to more complex settings involving many categorical, count, continuous and even functional observations.

Another interesting direction in terms of generalizations is to accommodate dependence in the observations; for example, one may collected multivariate categorical longitudinal data in which the same variables are measured repeatedly on the sample study subjects or the data may have a nested structure. Log linear and logistic regression-type models can be easily generalized to such settings, but clearly encounter computational challenges in large sparse settings. Potentially the simplex factor model of Bhattacharya and Dunson (2012) can be generalized to accommodate such dependence structures through the latent factors, with some challenges arising in terms of developing computationally efficient implementations and models that are both flexible and interpretable.



\subsection*{Acknowledgement}

This work was supported by Nakajima Foundation and grant number R01 ES017240 from the National Institute of Environmental Health Sciences (NIEHS) of the National Institutes of Heath (NIH).  The computational results are mainly generated using Ox (\cite{Doornik06}).


\appendix

\section{Proof of Lemma 1} \label{ap:2}
The expectation of cell probability is 
\begin{align*}
E\{\pi_{tc_1\cdots c_p}\} &= E\left\{ \sum_{h=1}^{\infty} \nu_{th} \prod_{j=1}^p \psi^{(j)}_{hc_j} \right\} = \sum_{h=1}^{\infty} \left[ E\{\nu_{th}\} \prod_{j=1}^p E\left\{\psi^{(j)}_{hc_j}\right\} \right], \\
&= \prod_{j=1}^p E\left\{\psi^{(j)}_{hc_j}\right\} \sum_{h=1}^{\infty} E\{\nu_{th}\} = \prod_{j=1}^p E\left\{\psi^{(j)}_{hc_j}\right\} = \prod_{j=1}^p \frac{a_{jc_j}}{\hat{a}_j}. \\
\end{align*}
The marginal distribution of $W_{th}$ can be expressed as $N(\mu/(1-\phi), \sigma^2_{\eta}/(1-\phi^2)+\sigma^2_{\varepsilon})$, independent of $t$ and $h$.  Hence, we set $\beta_{1} = E\{g(W_{th})\}$ and $\beta_{2} = E\{g^2(W_{th})\}$.
The second moment of cell probability is
\begin{align*}
E\{ \pi_{tc_1\cdots c_p}^2 \} &= E \left[ \left\{ \sum_{h=1}^{\infty} \nu_{th} \prod_{j=1}^p \psi^{(j)}_{hc_j} \right\} \left\{ \sum_{l=1}^{\infty} \nu_{tl} \prod_{j=1}^p \psi^{(j)}_{lc_j} \right\} \right], \\
&= \sum_{h=1}^{\infty} \sum_{l=1}^{\infty} E\{\nu_{th}\nu_{tl}\} E\left\{ \prod_{j=1}^p \psi^{(j)}_{hc_j} \psi^{(j)}_{lc_j}  \right\}, \\
&= \left[ \prod_{j=1}^p E \left\{ \left( \psi^{(j)}_{hc_j} \right)^2 \right\} - \prod_{j=1}^p E^2\left\{ \psi^{(j)}_{hc_j} \right\}  \right] \sum_{h=1}^{\infty} E\{ \nu_{th}^2 \} + \prod_{j=1}^p E^2\left\{ \psi^{(j)}_{hc_j} \right\} \sum_{h=1}^{\infty} \sum_{l=1}^{\infty} E\{\nu_{th}\nu_{tl}\}, \\
&= \left( \prod_{j=1}^p \frac{a_{jc_j}(a_{jc_j}+1)}{\hat{a}_j(\hat{a}_j+1)} - \prod_{j=1}^p \frac{a^2_{jc_j}}{\hat{a}^2_{j}} \right) \sum_{h=1}^{\infty} E\{ \nu_{th}^2 \} + \prod_{j=1}^p \frac{a^2_{jc_j}}{\hat{a}^2_j}, 
\end{align*}
\vspace{-5mm}
where
\begin{align*}
\sum_{h=1}^{\infty} E\{ \nu_{th}^2 \} &= \sum_{h=1}^{\infty} E \left[ g^2(W_{th}) \prod_{l<h} \{1 - g(W_{tl})\}^2 \right], \\
&= \sum_{h=1}^{\infty} \beta_2 \{ 1 -2\beta_1 + \beta_2 \}^{h-1},\\
&= \frac{\beta_2}{2\beta_1-\beta_2}.
\end{align*}
Hence,
\begin{align}
V\{\pi_{tc_1\cdots c_p}\} = \left( \prod_{j=1}^p \frac{a_{jc_j}(a_{jc_j}+1)}{\hat{a}_j(\hat{a}_j+1)} - \prod_{j=1}^p \frac{a^2_{jc_j}}{\hat{a}^2_{j}} \right)\left( \frac{\beta_2 }{2\beta_1-\beta_2} \right).
\label{eq:variance}
\end{align}
Similarly,
\begin{align*}
E&\{\pi_{tc_1\cdots c_p}\pi_{t+kc'_1\cdots c'_p}\} = E \left[ \left\{ \sum_{h=1}^{\infty} \nu_{th} \prod_{j=1}^p \psi^{(j)}_{hc_j} \right\} \left\{ \sum_{l=1}^{\infty} \nu_{t+kl} \prod_{i=1}^p \psi^{(i)}_{lc'_i} \right\} \right], \\
&= \left[ \prod_{j=1}^p E\left\{ \psi^{(j)}_{hc_j} \psi^{(j)}_{hc'_j} \right\} - \prod_{j=1}^p  E\left\{ \psi^{(j)}_{hc_j} \right\} E\left\{ \psi^{(j)}_{lc'_j} \right\}  \right] \sum_{h=1}^{\infty} E\{ \nu_{th} \nu_{t+kh} \}  
+ \prod_{j=1}^p  E\left\{ \psi^{(j)}_{hc_j} \right\} E\left\{ \psi^{(j)}_{lc'_j} \right\}, \\ 
&= \left( \prod_{j=1}^p \frac{a_{jc_j} \{ a_{jc'_j} + 1(c_j=c'_j) \}}{\hat{a}_j (\hat{a}_j + 1)} - \prod_{j=1}^p \frac{a_{jc_j} a_{jc'_j} }{\hat{a}^2_j} \right) \sum_{h=1}^{\infty} E\{ \nu_{th} \nu_{t+kh} \} + \prod_{j=1}^p \frac{a_{jc_j} a_{jc'_j} }{\hat{a}^2_j},
\end{align*}
where
\begin{align*}
E\{ \nu_{th} \nu_{t+kh} \}&= E \left\{ \left[ g(W_{th}) \prod_{l<h} \{1 - g(W_{tl})\} \right] \left[ g(W_{t+kh}) \prod_{l<h} \{1 - g(W_{t+kl})\} \right] \right\}, \\
&= E \left\{ g(W_{th}) g(W_{t+kh}) \right\} \prod_{l<h} E \left[ \{1 - g(W_{tl})\}\{1 - g(W_{t+kl})\} \right], \\
&= E \left\{ g(W_{th}) g(W_{t+kh}) \right\} \prod_{l<h} \left[ 1 - 2\beta_1 + E\{ g(W_{tl}) g(W_{t+kl}) \}  \right].
\end{align*}
From (\ref{eq:new4}) and (\ref{eq:new5}), $E \left\{ g(W_{th}) g(W_{t+kh}) \right\}$ can be expressed as 
\begin{align*}
E \left\{ g(W_{th}) g(W_{t+kh}) \right\} &= E \left\{ g(\alpha_{th}+\varepsilon_{th}) g(\alpha_{t+kh}+\varepsilon_{t+kh}) \right\}, \\
&= E \left\{ g\left(\alpha_{th} + \varepsilon_{th}\right) g\left(\frac{1-\phi^k}{1-\phi}\mu + \phi^k \alpha_{th} + \sum_{i=0}^{k-1} \phi^i w_{t+k-ih} + \varepsilon_{t+kh}\right) \right\}.
\end{align*}
Since $\alpha_{th}$, $\varepsilon_{th}$, $w_{t+k-ih}$ ($i=0,\ldots,k-1$) and $\varepsilon_{t+kh}$ are independent of one another and their distributions do not depend on $t$ or $h$, hence $\gamma_k\equiv E \left\{ g(W_{th}) g(W_{t+kh}) \right\}$ is dependent on time difference $k$ but independent of time $t$.

In addition,
\begin{align*}
\sum_{h=1}^{\infty} E\{ \nu_{th} \nu_{t+kh} \} &= \sum_{h=1}^{\infty} \gamma_k \prod_{l<h} \left\{ 1 - 2\beta_1 + \gamma_k \right\}, \\
&= \frac{\gamma_k}{2\beta_1 - \gamma_k}.
\end{align*}
Hence,
\begin{align*}
Cov\{\pi_{tc_1\cdots c_p}, \pi_{t+k c'_1\cdots c'_p}\} = \left( \prod_{j=1}^p \frac{a_{jc_j} \{ a_{jc'_j} + 1(c_j=c'_j) \}}{\hat{a}_j (\hat{a}_j + 1)} - \prod_{j=1}^p \frac{a_{jc_j} a_{jc'_j} }{\hat{a}^2_j} \right)\left( \frac{\gamma_k}{ 2\beta_1-\gamma_k } \right).
\end{align*}
Since $\beta_2 /(2\beta_1-\beta_2)>0$, $\gamma_k/(2\beta_1-\gamma_k)>0$ and (\ref{eq:variance}), cell probabilities with $c_j= c'_j$ for all $j$ have positive covariance and, on the other hand, those with $c_j\neq c'_j$ for all $j$ have negative covariance. 

In a case where $a_{j1}=\cdots=a_{jc_j}=a$, the variance and covariance are expressed as
\begin{align*}
V\{\pi_{tc_1\cdots c_p}\} &= \left( \prod_{j=1}^p \frac{1+1/a}{d^2_j+d_j/a} - \prod_{j=1}^p \frac{1}{d^2_j} \right)\left( \frac{\beta_2 }{2\beta_1-\beta_2} \right), \\
Cov\{\pi_{tc_1\cdots c_p}, \pi_{t+k c'_1\cdots c'_p}\} &= \left( \prod_{j=1}^p \frac{1+1(c_j=c'_j)/a}{d_j^2 + d_j/a } - \prod_{j=1}^p \frac{1}{d_j^2} \right) \left( \frac{\gamma_k}{ 2\beta_1-\gamma_k } \right). 
\end{align*}
Hence, $V\{\pi_{tc_1\cdots c_p}\}\rightarrow 0$ and $Cov\{\pi_{tc_1\cdots c_p}, \pi_{t+k c'_1\cdots c'_p}\}\rightarrow 0$ as $a\rightarrow \infty$.

\section{Proof of Lemma 2} \label{ap:1}

To prove $\sum_{h=1}^{\infty}\nu_{th}=1$ a.s., it is enough to show $\sum_{h=1}^{\infty}E\{\log(1-g(W_{th})\}=-\infty$ (\cite{IshwaranJames01}).  $g$ is a non-negative monotone increasing link function: $\Re \to (0,1)$, therefore $0<\beta_1 =E\{g(W_{th})\}<1$.  Then, using Jensen's inequality,
\begin{align*}
E[\log\{1-g(W_{th})\}] \leq \log[1 - E\{g(W_{th})\}] = \log(1 - \beta_1) < 0. 
\end{align*}
Therefore, $\sum_{h=1}^{\infty}E\{\log(1-g(W_{th})\}=-\infty$ at each time point.

\section{Proof of theorem} \label{ap:3}

The proposed prior probability assigned to $\mathcal{N}_{\epsilon}(\bm{\pi}^0)$ can be expressed as
\begin{align*}
\mathcal{Q} \left\{ \mathcal{N}_{\epsilon}(\bm{\pi}^0) \right\} = \int 1( \| \bm{\pi} - \bm{\pi}^0 \| < \epsilon ) d \mathcal{Q}(\bm{\nu}_t, \bm{\psi}_{h}^{(j)}, t\in\{1,\ldots,T\}, h=1,\ldots,\infty, j=1,\ldots,p).
\end{align*} 
where $\bm{\nu}_t$ is a probability vector induced by the proposed stick breaking process and we use the $L_1$ distance  
\begin{align*}
\| \bm{\pi} - \bm{\pi}^0 \| = \sum_{t=1}^T p_t \sum_{c_1=1}^{d_1} \cdots \sum_{c_p=1}^{d_p} | \pi_{tc_1\cdots c_p} - \pi_{tc_1\cdots c_p}^0 |,
\end{align*} 
where $p_t$ is a probability mass function for time $t\in\{1,\ldots,T\}$.  


For any $\bm{\pi}^0 \in \Pi$, each component in $\bm{\pi}^0$ can be expressed as
\begin{align*}
\bm{\pi}^0_t &= \sum_{h=1}^{k_t} \nu^0_{th} \Psi_{th}, \ \ \ \ \Psi_{th} = \bm{\psi}_{th}^{(1)} \otimes \cdots \otimes \bm{\psi}_{th}^{(p)},
\end{align*}
where $k_t\in \mathbb{N}$, $\bm{\nu}^0_t = (\nu^0_{t1},\ldots,\nu^0_{tk_t})'$ is a probability vector, $\Psi_{th}\in \Pi_{d_1\cdots d_p}$ and $\bm{\psi}_{th}^{(j)}=(\psi^{(j)}_{th1},\ldots,\psi^{(j)}_{thd_j})'$ is a $d_j\times 1$ probability vector.  We define $k^+_0=0$ and $k^{+}_t=\sum_{i=1}^{t}k_i$ for $t=1,\ldots,T$. Then, we construct $\bm{\pi}=\{\bm{\pi}_t,t\in\{1,\ldots,T\}\} \in \Pi$ induced by the proposed prior such that the component with the index $h$ in $\bm{\pi}^0_t$ is approximated by the component with the index $k^+_{t-1}+h$ in $\bm{\pi}_t$.  Let $\tilde{\bm{\nu}}_t=(\tilde{\nu}_{t1},\tilde{\nu}_{t2},\ldots)'$ be a probability vector where $\tilde{\nu}_{tm}=\nu^0_{tm-k^+_{t-1}}$ for $k^+_{t-1} <  m\leq k^+_{t}$ and $\tilde{\nu}_{tm}=0$ otherwise, i.e., $\tilde{\nu}_{tk^+_{t-1}+h}=\nu^0_{th}$ for $1 \leq  h\leq k_{t}$.  For any $\epsilon$, we define a set $D(\bm{\pi}^0,\epsilon)\subset \Pi$
 such that for any $\bm{\pi} \in D(\bm{\pi}^0,\epsilon)$, each $\bm{\pi}_t$ can be expressed as (\ref{eq:new1}) satisfying $\bm{\nu}\in\mathcal{N}_{\epsilon'}(\tilde{\bm{\nu}})$, where $\bm{\nu}= \{\bm{\nu}_t,t\in\{1,\ldots,T\}\}$, $\tilde{\bm{\nu}}= \{\tilde{\bm{\nu}}_t,t\in\{1,\ldots,T\}\}$ and $\epsilon'=\epsilon/2\prod_{j=1}^{p}d_j$, and $\bm{\psi}_{k^+_{t-1}+h}^{(j)}\in \mathcal{N}_{\epsilon''}\left(\bm{\psi}_{th}^{(j)}\right)$ for $h=1,\ldots,k_t$ and $t=1,\ldots,T$ where $\epsilon''=\epsilon/2\sum_t p_t k_t  p \prod_{j}d_j$.

We consider the intervals $(a_{th}, b_{th})$ in the real line for $W_{th}$ in the proposed prior for $h=1,\ldots,k^+_{t}$ and $t=1,\ldots,T$ where 
\begin{equation*}
a_{th} = \begin{cases}
             g^{-1}\{\max(\tilde{\nu}_{th}-\tilde{\epsilon}, 0)\}, & (h=1),\\
             g^{-1}\left\{ \frac{\max(\tilde{\nu}_{th}-\tilde{\epsilon}, 0)}{\prod_{l<h} \{1-g(W_{tl})\} } \right\}, & (h=2,\ldots, k^+_{t}),
       \end{cases}       
b_{th} = \begin{cases}
             g^{-1}\{\tilde{\nu}_{th} + \tilde{\epsilon}\}, & (h=1),\\
             g^{-1}\left\{ \frac{\tilde{\nu}_{th}+\tilde{\epsilon}}{\prod_{l<h} \{1-g(W_{tl})\} } \right\}, & (h=2,\ldots, k^+_{t}),
       \end{cases}
\end{equation*}
where $\tilde{\epsilon}=\epsilon'/2\sum_t p_t k_t^+$.  In this case, it is straightforward to check $| \nu_{th} - \tilde{\nu}_{th} | < \tilde{\epsilon}$ for $h=1,\ldots,k^+_{t}$ and the proposed prior assigns positive probability to these intervals.  Then, the distance between $\bm{\nu}$ and $\tilde{\bm{\nu}}$ is 
\begin{align}
\| \bm{\nu} - \tilde{\bm{\nu}} \| &= \sum_{t=1}^T p_t \sum_{h=1}^{\infty} | \nu_{th} - \tilde{\nu}_{th} |, \nonumber \\
&= \sum_{t=1}^T p_t \sum_{h=1}^{k_t^+} | \nu_{th} - \tilde{\nu}_{th} | + \sum_{t=1}^T p_t \sum_{h>k_t^+} \nu_{th}, \label{eq:nu} \\
&< 2 \tilde{\epsilon} \sum_{t=1}^T p_t k_t^+ = \epsilon'. \nonumber
\end{align}
For the second component in (\ref{eq:nu}), $\sum_{h>k_t^+} \nu_{th} <  k_t^+ \tilde{\epsilon}$ because $\nu_{th}>\tilde{\nu}_{th}-\tilde{\epsilon}$ for $h=1,\ldots,k_t^+$ and $\sum_{h=1}^{k_t^+} \nu_{th} > 1 -k_t^+ \tilde{\epsilon}$.  In addition, it is straightforward to show that the proposed prior assigns positive probability to $\mathcal{N}_{\epsilon''}\left(\bm{\psi}_{th}^{(j)}\right)$.  Therefore, since $D(\bm{\pi}^0,\epsilon)$ contains such case, $\mathcal{Q}\{D(\bm{\pi}^0,\epsilon)\}>0$.

For any $\bm{\pi}\in D(\bm{\pi}^0,\epsilon)$,
\begin{align*}
\| \bm{\pi} - \bm{\pi}^0 \| &= \sum_{t=1}^T p_t \sum_{c_1=1}^{d_1} \cdots \sum_{c_p=1}^{d_p} | \pi_{tc_1\cdots c_p} - \pi_{tc_1\cdots c_p}^0 |, \\
&= \sum_{t=1}^T p_t \sum_{c_1=1}^{d_1} \cdots \sum_{c_p=1}^{d_p} \left| \sum_{h=1}^{\infty} \nu_{th}\prod_{j=1}^p \psi_{hc_j}^{(j)} - \sum_{l=1}^{k_t} \nu^0_{tl}\prod_{j=1}^p \psi_{tlc_j}^{(j)} \right|, \\
&= \sum_{t=1}^T p_t \sum_{c_1=1}^{d_1} \cdots \sum_{c_p=1}^{d_p} \left| \sum_{h=1}^{k_t} \left( \nu_{tk^+_{t-1}+h}\prod_{j=1}^p \psi_{k^+_{t-1}+hc_j}^{(j)} - \nu^0_{th}\prod_{j=1}^p \psi_{thc_j}^{(j)} \right) + \sum_{l\leq k^+_{t-1}, k^+_{t}<l} \nu_{tl}\prod_{j=1}^p \psi_{lc_j}^{(j)} \right|, \\
&\leq \sum_{t=1}^T p_t \sum_{c_1=1}^{d_1} \cdots \sum_{c_p=1}^{d_p} \left( \sum_{h=1}^{k_t} \left| \nu_{tk^+_{t-1}+h}\prod_{j=1}^p \psi_{k^+_{t-1}+hc_j}^{(j)} - \nu^0_{th}\prod_{j=1}^p \psi_{thc_j}^{(j)} \right| + \sum_{l\leq k^+_{t-1}, k^+_{t}<l} \nu_{tl} \right), \\
&\leq  \sum_{t=1}^T p_t \sum_{c_1=1}^{d_1} \cdots \sum_{c_p=1}^{d_p} \left( \sum_{h=1}^{k_t} \left| \nu_{tk^+_{t-1}+h} - \nu^0_{th} \right| + \sum_{l=1}^{k_t} \sum_{j=1}^p \left| \psi_{k^+_{t-1}+lc_j}^{(j)} - \psi_{tlc_j}^{(j)} \right| + \sum_{l\leq k^+_{t-1}, k^+_{t}<l} \nu_{tl} \right), \\
&= \sum_{c_1=1}^{d_1} \cdots \sum_{c_p=1}^{d_p} \sum_{t=1}^T p_t \sum_{h=1}^{\infty} | \nu_{th} - \tilde{\nu}_{th} |  + \sum_{t=1}^T p_t \sum_{l=1}^{k_t} \sum_{j=1}^p \sum_{c_1=1}^{d_1} \cdots \sum_{c_p=1}^{d_p} \left| \psi_{k^+_{t-1}+lc_j}^{(j)} - \psi_{tlc_j}^{(j)} \right|, \\ 
&< \prod_{j=1}^{p}d_j \epsilon' + \sum_{t=1}^{T} p_t k_t  p \prod_{j=1}^p d_j \epsilon'',  \\
&= \frac{\epsilon}{2} + \frac{\epsilon}{2} = \epsilon.
\end{align*}
Therefore $\bm{\pi}\in \mathcal{N}_{\epsilon}(\bm{\pi}^0)$ and $D(\bm{\pi}^0,\epsilon) \subset  \mathcal{N}_{\epsilon}(\bm{\pi}^0)$. Hence, $\mathcal{Q}\{ \mathcal{N}_{\epsilon}(\bm{\pi}^0) \} > 0$.

\section{Figures}

\bibliographystyle{chicago}
\bibliography{tct}

\newpage

\section{Supplemental materials}

\begin{table}[H]
\centering
\small
\begin{tabular}{clr}

\hline
\multicolumn{1}{c}{No.}		& \multicolumn{1}{l}{Categorical variable (Name in GSS)}	&	\multicolumn{1}{r}{\scalebox{0.7}[1]{\# of categories}} \\
\hline 
1 & Age group* (AGE) & 8 \\
2 & Sex (SEX) & 2\\
3 & Race (RACE) & 3 \\  
4 & Religious preference** (RELIG) & 5\\
5 & Region (REGION) & 9 \\
6 & Attitude toward abortion (ABANY) & 2 \\
7 & Should Govetnment help pay for medical care? (HELPSICK) & 5 \\
8 & Highest degree (DEGREE) & 5 \\
9 & Political party affiliation (PARTYID) & 8 \\
10 & Current marital status (MARITAL) & 5 \\
11 & Astrological sign (ZODIAC) & 12 \\
12 & Confidence in banks and financial institutions (CONFINAN) & 3 \\
13 & Confidence in U.S. Supreme Court (CONJUDGE) & 3 \\
14 & Think of self as liberal or conservative (POLVIEWS) & 7 \\
15 & Belief in life after death (POSTLIFE) & 2 \\
16 & Attitude toward homosexual sex relations (HOMOSEX) & 5 \\
17 & Have gun in home (OWNGUN) & 2 \\
18 & Subjective class identification (CLASS) & 4 \\
19 & Should Marijuana be made legal (GRASS) & 2 \\
20 & Total family income (INCOME) & 12 \\
21 & Favor or oppose death penalty for murder (CAPPUN) & 2 \\
22 & \scalebox{0.83}[1]{Attitude toward spending money on space exploration program (NATSPAC)} & 3 \\ 
23 & \scalebox{0.83}[1]{Attitude toward spending money on improving and protecting environment (NATENVIR)} & 3 \\
24 & \scalebox{0.83}[1]{Attitude toward spending money on improving and protecting the nations's health (NATHEAL)} & 3 \\
25 & \scalebox{0.83}[1]{Attitude toward spending money on halting the rising crime rate (NATCRIME)} & 3 \\
26 & \scalebox{0.83}[1]{Attitude toward spending money on dealing with drug addiction (NATDRUG)} & 3 \\
27 & \scalebox{0.83}[1]{Attitude toward spending money on improving the nation's education system (NATEDUC)} & 3 \\
28 & \scalebox{0.83}[1]{Attitude toward spending money on the military, armaments and defense (NATARMS)} & 3 \\
29 & \scalebox{0.83}[1]{Attitude toward spending money on foreigh aid (NATAID)} & 3 \\
\hline
\end{tabular}
\begin{minipage}{15cm}
{\footnotesize *The category of Age group is different from the original one: 1. 18 or 19 years old, 2. 20s, 3. 30s, 4. 40s, 5. 50s, 6. 60s, 7. 70s, 8. more than 80 years old. \\
**The category of Religious preference is different from the original one: 1. Protestant, 2. Catholic, 3. Jewish, 4. None, 5. Others.}
\end{minipage}
\normalsize
\caption{List of categorical variables.}
\end{table}

\begin{figure}[H]
  \begin{center}
    \subfigure{\includegraphics[scale=0.7]{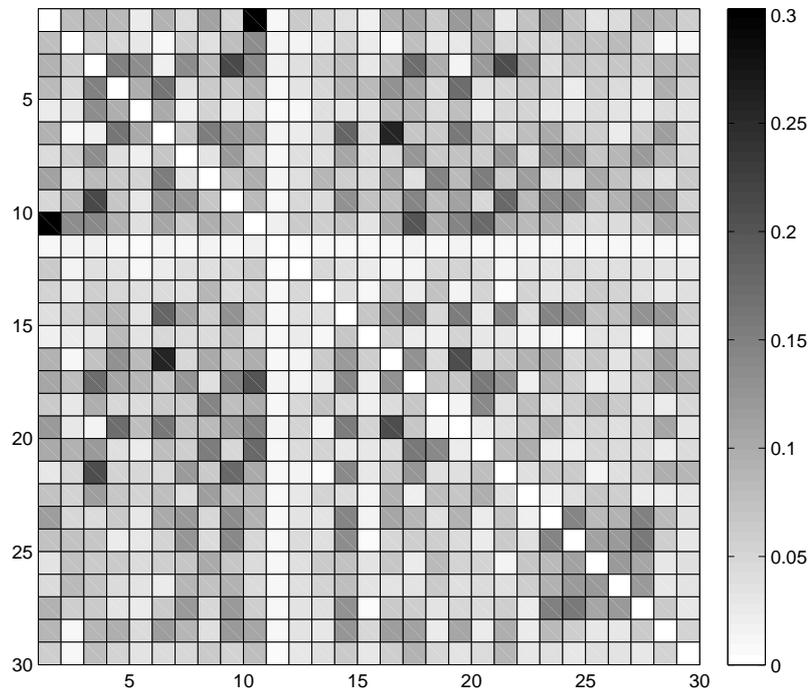}}
	\subfigure{\includegraphics[scale=0.7]{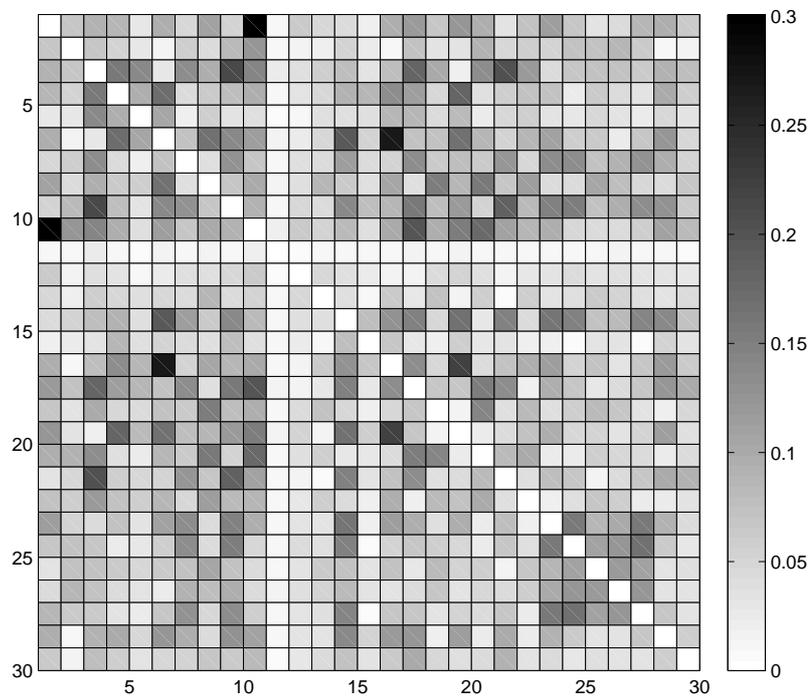}}
  \end{center}
\caption{Posterior means of $\rho_{tjj'}$ in 1994 (above) and 1996 (below).}
\end{figure}

\begin{figure}[H]
  \begin{center}
    \subfigure{\includegraphics[scale=0.7]{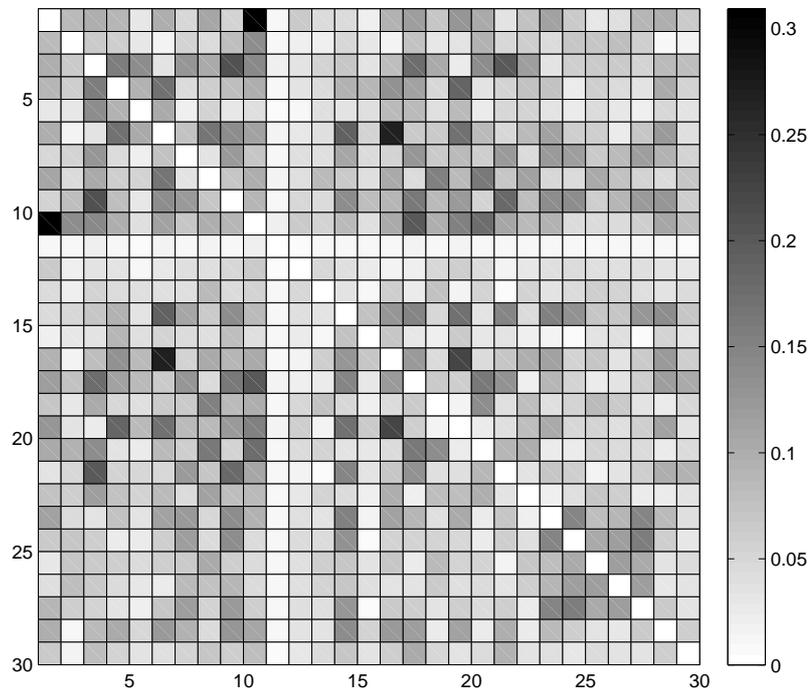}}
	\subfigure{\includegraphics[scale=0.7]{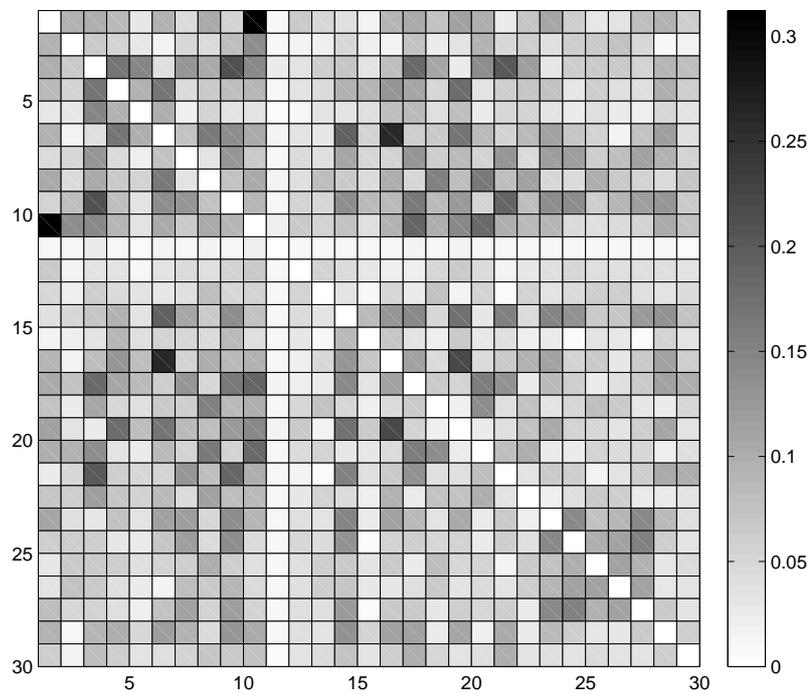}}
  \end{center}
\caption{Posterior means of $\rho_{tjj'}$ in 1998 (above) and 2000 (below).}
\end{figure}

\begin{figure}[H]
  \begin{center}
    \subfigure{\includegraphics[scale=0.7]{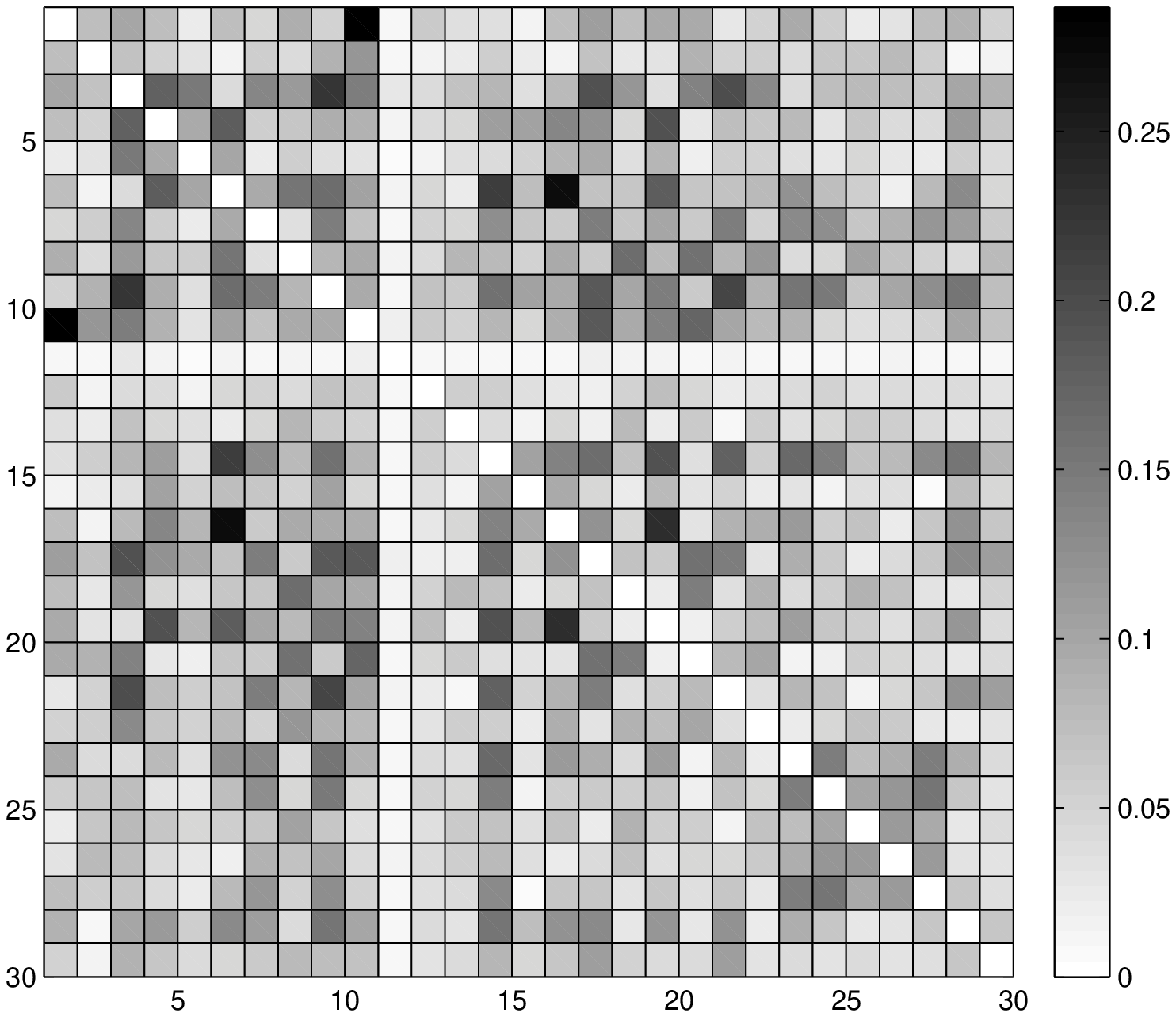}}
	\subfigure{\includegraphics[scale=0.7]{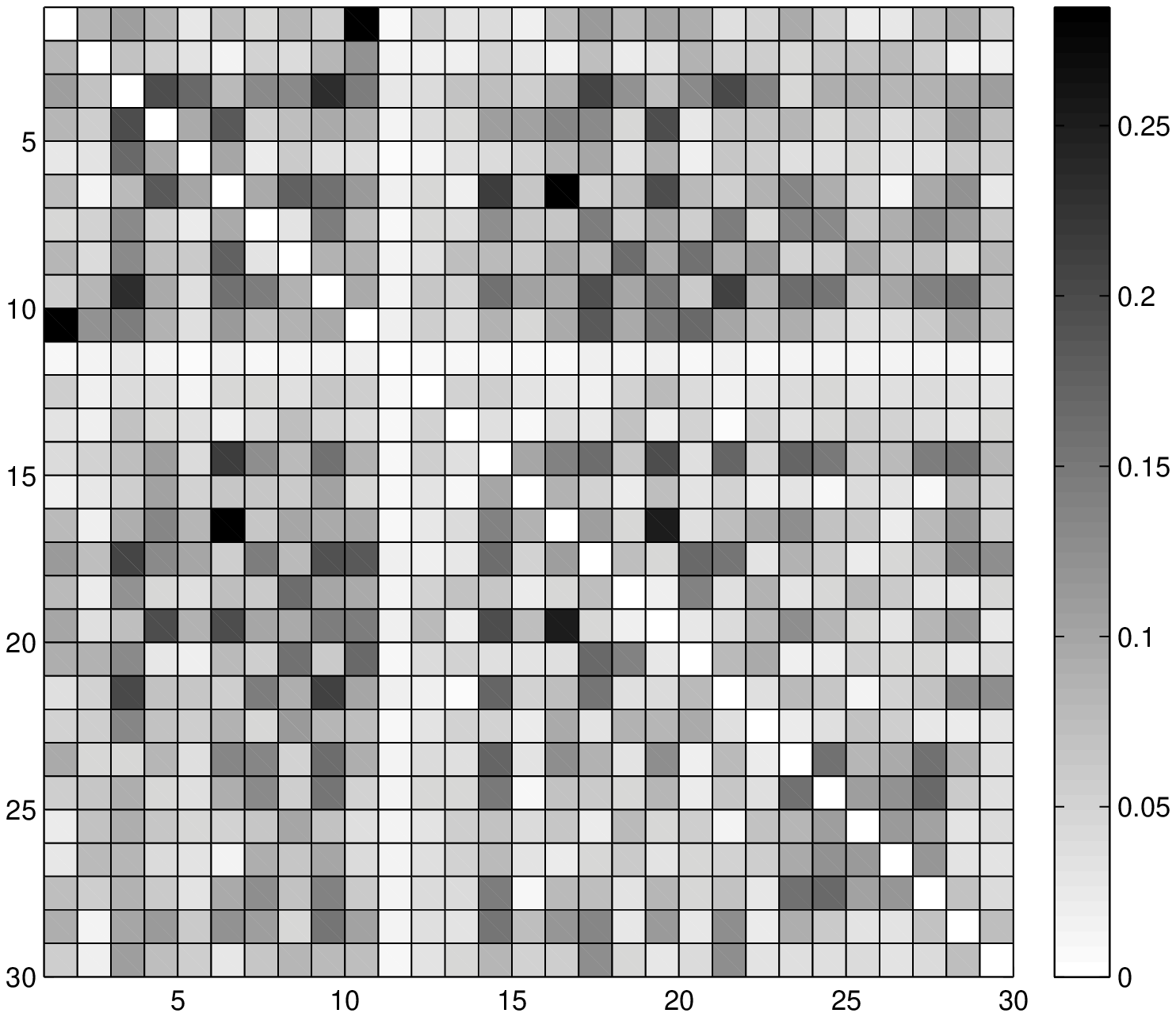}}
  \end{center}
\caption{Posterior means of $\rho_{tjj'}$ in 2004 (above) and 2006 (below).}
\end{figure}

\begin{figure}[H]
  \begin{center}
    \includegraphics[scale=0.7]{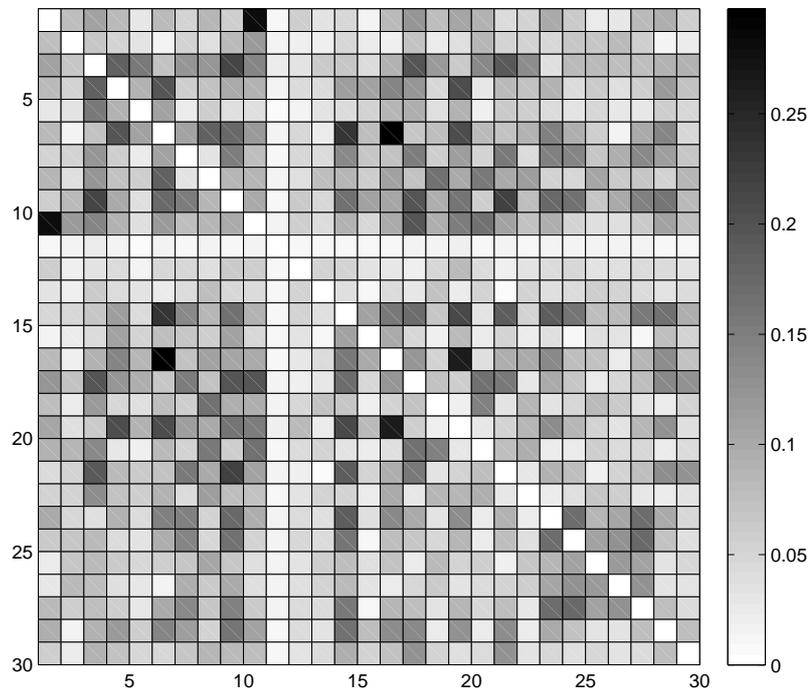}
  \end{center}
\caption{Posterior means of $\rho_{tjj'}$ in 2008.}
\end{figure}

\end{document}